\documentclass[12pt]{article}
\usepackage{amssymb}
\usepackage{amsmath}

\newtheorem{theorem}{Theorem}

\newtheorem{corollary}{Corollary}

\begin{document}

\author{{Roustam Zalaletdinov$^{(1,2)}$ and Alan Coley$^{(1)}$} \\ [5mm]
\emph{$^{(1)}$Department of Mathematics and Statistics, Dalhousie University}\\
\emph{Chase Building, Halifax, Nova Scotia, Canada B3H 3J5} \\ [2mm]
\emph{$^{(2)}$Department of Theoretical Physics, Institute of Nuclear Physics} \\
\emph{Uzbek Academy of Sciences, Ulugbek, Tashkent 702132, Uzbekistan, CIS} \\
}
\title{{\LARGE \textbf{Averaging out Inhomogeneous Newtonian Cosmologies:
III. The Averaged Navier-Stokes-Poisson Equations }}}
\date{}
\maketitle

\begin{abstract}
The basic concepts and hypotheses of Newtonian Cosmology necessary for a
consistent treatment of the averaged cosmological dynamics are formulated
and discussed in details. The space-time, space, time and ensemble averages
for the cosmological fluid fields are defined and analyzed with a special
attention paid to their analytic properties. It is shown that all averaging
procedures require an arrangement for a standard measurement device with the
same measurement time interval and the same space region determined by the
measurement device resolution to be prescribed to each position and each
moment of time throughout a cosmological fluid configuration. The formulae
for averaging out the partial derivatives of the averaged cosmological fluid
fields and the main formula for averaging out the material derivatives have
been proved. The full system of the averaged Navier-Stokes-Poisson equations
in terms of the fluid kinematic quantities is derived.
\end{abstract}

\section{Introduction}

In order to describe the dynamics of inhomogeneous and anisotropic Newtonian
universes one needs to establish a set of averaged Navier-Stokes-Poisson
equations in kinematic quantities. To approach this problem the basic
concepts and hypotheses of Newtonian Cosmology necessary for a consistent
treatment of averaged cosmological dynamics are formulated and discussed in
details. The space-time, space, time and ensemble averages for the
cosmological fluid fields are defined and analyzed with a special attention
paid for their analytic properties. It is shown that all averaging
procedures require an arrangement for a standard measurement device with the
same measurement time interval and the same space region determined by the
measurement device resolution to be prescribed to each position and each
moment of time throughout a cosmological fluid configuration. The formulae
for averaging out the partial derivatives of the average fluid fields and
the main formula for averaging out the material derivatives have been
proved. The full set of the averaged Navier-Stokes-Poisson equations in
terms of the fluid kinematic quantities is derived.

The structure of the paper is as follows. The Averaging problem in general
relativity which has originated in cosmology when it has been realized that
a modification of the Einstein equations by smoothing them out in some
appropriate sense should be established for a description of the large-scale
structure of the Universe is discussed in Chapter 2. It is shown that
Newtonian cosmology also has the Averaging problem, since due to the
nonlinearity of the Navier-Stokes-Poisson equations one needs to prove that
the averaged Navier-Stokes-Poisson equations have a solution for a
homogeneous and isotropic Newtonian universe which can be obtained upon
averaging out a locally inhomogeneous and/or anisotropic cosmological fluid
configuration representing evolving matter structures. Chapter 3 shows that
an averaged description of a Newtonian universe means, from the physical
point of view, that the dynamics of the Newtonian universe is written
directly in terms of observationally relevant measurable quantities. The
Newtonian cosmological macroscopic hypothesis which assumes the existence of
two essentially distinct types of temporal and spatial scales, namely, the
local scales of cosmological fluid fluctuations and cosmological mean
scales, is formulated and discussed in details in Chapter 4. A
classification of Newtonian cosmologies is given, in particular, on the
basis of this hypothesis. Chapter 5 gives the definitions of the space-time
averages of cosmological fluid fields and their correlations and discusses
how one can construct an average fluid field by arranging a covering of the
Newtonian space-time by a standard measurement device at every moment of
time and at every position throughout a cosmological fluid configuration.
The Reynolds conditions necessary for any kinds of averages are formulated
and proved for the space-time averages in Chapter 6. Properties and
conditions on the space-time correlation functions are proved in Chapter 7.
Chapter 8 is devoted to the definition and analysis of time averages and
time correlation functions. It is shown, in particular, that time averages
over an infinite time interval does not depend on time. Chapter 9 is devoted
to the definition and analysis of space averages and space correlation
functions. It is shown that space averages over the whole space are constant
with respect to the space position. The formula for the space-time averaging
out of the material derivatives is proved in Chapter 10 and the form of the
space-time averaged Navier-Stokes operator is derived. In Chapter 11 a
possibility to use the Reynolds transport theorem for a definition of space
averages is analyzed and it is shown that one cannot take this approach for
the proper treatment of space averages. In Chapter 12 the ensemble averages
and ensemble correlation functions are defined and the statistical treatment
of cosmological fluid fields is discussed. The Hypothesis of the statistical
nature of Newtonian universes is also formulated here. Properties and
conditions on the ensemble averages and ensemble correlation functions, such
as the Reynolds conditions, the formula for the ensemble averaging out of
the material derivatives and the form the ensemble averaged Navier-Stokes
operator,\ are proved in Chapter 13. The Ergodic hypothesis of Newtonian
cosmology is discussed and formulated in Chapter 14. The full set of the
averaged Navier-Stokes-Poisson equations in terms of the fluid kinematic
quantities valid for the space-time and ensemble averages is derived in
Chapter 15. It is pointed out that the system must be supplemented by a
system of the equations for the single-point second order moments of fluid
fields present in the averaged Navier-Stokes-Poisson equations.

The formulae and Sections from the papers I and II \cite{Zala-AIC1:2002},
\cite{Zala-AIC2:2002} will be referred to as (I-XX) and I-X, and (II-XX) and
II-X, correspondingly.

Conventions and notations are as follows. All functions $f=f(x^{i},t)$ are
defined on 3-dimensional Euclidean space $E^{3}$ in the Cartesian
coordinates $\{x^{i}\}$ with Latin space indices $i,j,k,...$ running from 1
to 3, and $t$ is the time variable. The Levi-Civita symbol $\varepsilon
_{ijk}$ is defined as $\varepsilon _{123}=+1$ and $\varepsilon ^{123}=+1$
and $\delta ^{ij}$, $\delta _{j}^{i}$ and $\delta _{ij}$ are the Kronecker
symbols. The symmetrization of indices of a tensor $
T_{jk}^{i}=T_{jk}^{i}(x^{l},t)$ is denoted by round brackets, $T_{(jk)}^{i}=
\frac{1}{2}\left( T_{jk}^{i}+T_{kj}^{i}\right) $, and the antisymmetrization
by square brackets, $T_{[jk]}^{i}=\frac{1}{2}\left(
T_{jk}^{i}-T_{kj}^{i}\right) $. A partial derivative of $T_{jk}^{i}$ with
respect to a spatial coordinate $x^{i}$ or time $t$ is denoted either by
comma, or by the standard calculus notation, $T_{jk,t}^{i}=\partial
T_{jk}^{i}/\partial t$ and $T_{jk,l}^{i}=\partial T_{jk}^{i}/\partial x^{l}$
, the fluid velocity is $u^{i}=u^{i}(x^{j},t)$,\ and the material (total)
derivative is $\dot{T}_{jk}^{i}\equiv dT_{jk}^{i}/dt=\partial
T_{jk}^{i}/\partial t+u^{l}\partial T_{jk}^{i}/\partial
x^{l}=T_{jk,t}^{i}+u^{l}T_{jk,l}^{i}$. Average values are shown by either
the angular brackets, or by a bar over a tensor, $\langle T_{jk}^{i}\rangle
\equiv \overline{T}_{jk}^{i}$ with the corresponding subscripts to
distinguish between space-time, time, space and ensemble averages. The
Newton gravitational constant, the velocity of light and the Newtonian
cosmological constant are $G$, $c$ and $\Lambda $, respectively.

\section{The Averaging Problem in Newtonian Cosmology}

\label{*apnc}

Our Universe on the largest scales is assumed to be homogeneous and
isotropic though locally, at the scales of stars, galaxies and even clusters
of galaxies, it is inhomogeneous because of the matter condensation into
these discrete structures. Observational data definitely shows a high degree
of isotropy which is supported specifically by measurements of the
temperature of the cosmic background radiation. The large-scale spatial
homogeneity of the Universe up to scale of order 100-300 Mpc, however, still
remains an issue which cannot be considered as undoubtedly proved on the
basis of cosmological observations. In addition to uncertainties in the
measured data, difficulties in proving this fundamental hypothesis originate
also in the way how this data are treated. In particular, some theoretical
models for analysis of observational data are based on an assumption about
such a homogeneity \cite{Elli:2000}. Therefore, the general relativistic
Friedmann-Lemaitre-Robertson-Walker (FLRW) cosmological models \cite
{Frie:1922}-\cite{KSHM:1980} which are homogeneous and isotropic have not
been proved as yet to be precisely realistic to represent the large-scale
structure of our Universe even for the present epoch. The same is valid for
the Newtonian cosmological models, see Section II-9.

The observations of the large-scale characteristics of cosmological matter
and radiation do measure their averaged values over space regions and time
intervals. From physical point of view, one can think of the large-scale
cosmological models as representing an averaged picture of the Universe
which is highly inhomogeneous on the smaller scales characteristic of
typical matter structures. From the mathematical point of view one can think
of a cosmological model which is inhomogeneous on some scales to mimic local
gravitating matter condensations and which, at the same time, is homogeneous
on a large scale in some averaged sense, that is, after application of a
smoothing procedure to the physical fields and their governing equations.
Such a cosmological model would pretend to be more adequate to describe the
local dynamics of the real locally inhomogeneous Universe and its
large-scale homogeneous and/or isotropic dynamics. Then the Newtonian
cosmological principle, see Section II-6, is related to the large-scale
space-time structure of a locally inhomogeneous universe averaged out over
the space regions of a size much larger than a typical length of
inhomogeneities, say, the curvature radius of local matter structures, and
much less than a typical large-scale cosmological length, say, the curvature
radius of the Universe due to the mean matter density, and the time
intervals much larger than a typical time of change of local cosmological
variables governing the dynamics of the matter structures and much less than
a typical time of change of cosmological variables governing the evolution
of the Universe as a whole. The Averaging problem in the general
relativistic cosmology calls for a development of such a theoretical
physically reasonable and mathematically rigorous framework where one can
formulate a realistic inhomogeneous cosmological model which is homogeneous
and/or isotropic on average. The first attempt to carry out a space-time
volume averaging out the Einstein equations had been made by Shirokov and
Fisher \cite{Shir-Fish:1962}. The Averaging problem has been first
apparently formulated by Ellis \cite{Elli:1984} in the framework of general
relativistic cosmology (see \cite{Kras:1997} for a comprehensive review and
references). The first covariant approach for averaging out the Einstein
equations over space-time regions has been developed by Zalaletdinov \cite
{Zala:1992}-\cite{Zala:1997} in the framework of macroscopic gravity, a
classical macroscopic theory of gravity with the space-time volume averaged
(macroscopic) Einstein equations together with equations for the
gravitational correlation functions governing the dynamics of continuous
macroscopic matter distributions. The Einstein equations are considered as
the microscopic field equations suitable for description of point-like
matter distributions \cite{Zala:1997}, \cite{Tava-Zala:1998}. The
gravitational correlations have been shown to be very important for the
dynamics of the averaged (macroscopic) gravitational field, in particular,
when all correlations are assumed to vanish the macroscopic Einstein
equations become the Einstein equations of general relativity.

Thus, according to the Averaging problem a modification of the Einstein
equations by smoothing them out in some appropriate sense should be
established for a description of the large-scale structure of the Universe
and an adequate large-scale macroscopic cosmological model should be found
as a solution to these equations. The Einstein equations of general
relativity then have to describe the local dynamics of the Universe by a
suitable cosmological model. Upon an appropriate averaging of this
cosmological model one must obtain the macroscopic cosmological model which
is homogeneous and/or isotopic on some large cosmological scale.

The same problem exists for Newtonian cosmology. A distribution of stars,
galaxies and clusters of galaxies in the real Universe can be represented by
a locally inhomogeneous distribution of the cosmological fluid which is
governed either by the Navier-Stokes-Poisson equations (II-4), (II-8),
(II-9) and (II-10), or the system of Navier-Stokes-Poisson equations in
terms of kinematic quantities (II-4), (II-8), (II-12), (II-51)-(II-56),
(II-59)-(II-61), (I-64)-(II-66), (II-69), (II-28) and (II-70). Assuming the
Newtonian cosmological principle, see Section II-6, valid for the
large-scale structure of a Newtonian universe, one questions now whether or
not the suitably averaged Navier-Stokes-Poisson equations or the
Navier-Stokes-Poisson equations in terms of kinematic quantities have a
solution for an homogeneous and isotropic Newtonian universe such as this
large-scale homogeneous and isotropic cosmological fluid configuration can
be obtained upon averaging out the locally inhomogeneous and/or anisotropic
distribution of a cosmological fluid configuration representing evolving
matter structures.\vspace{0.2cm}

\noindent \textbf{The Newtonian averaging problem}\textsc{\hspace{0.2cm}}
\emph{Given a locally inhomogeneous and/or anisotro- pic distribution of a
cosmological fluid configuration representing the evolving cosmological
matter structures as a solution to the Navier-Stokes-Poisson equations
(II-4), (II-8), (II-9) and (II-10), if there exists a homogeneous and/or
isotropic Newtonian universe as a solution to the suitably averaged
Navier-Stokes-Poisson equations such as the homogeneous and isotropic
cosmological fluid configuration can be obtained upon averaging out this
locally inhomogeneous and/or anisotropic cosmological fluid configuration.
Then this locally inhomogeneous and/or anisotropic Newtonian universe will
satisfy the Newtonian cosmological principle for its large-scale structure. }
\vspace{0.2cm}

The importance of study of this difficult challenging problem is evident for
both Newtonian and general relativistic cosmologies. At the very fundamental
level, the Averaging problem questions the validity of the Newtonian
cosmological principle, see Section II-6, as a selection principle for a
realistic locally inhomogeneous and/or anisotropic model of the Universe to
be globally homogeneous and isotropic on its largest scales. From the
mathematical point of view, in general relativity due to the nonlinearity of
the Einstein equations one cannot expect this is to be true for an arbitrary
inhomogeneous cosmological model. Taking an average of the Einstein field
operator for the metric tensor serving as the gravitational potential in
general relativity does not commute with taking an average of the metric
tensor as it has been first pointed out by Shirokov and Fisher \cite
{Shir-Fish:1962} (see also \cite{Elli:1984}-\cite{Zala:1992}). In fluid
mechanics the same argument is valid for the Navier-Stokes-Poisson equations
as it has been known after Reynolds \cite{Reyn:1894}. Reynolds was the first
to realize that the nonlinearity of the Navier-Stokes equation (I-84)
actually contains a hidden key inside it to the understanding of fluid
motion through definition of the correlation functions of fluid velocity and
density for analytical study of the dynamics of evolving fluids by means of
a suitable time and/or space averaging procedure . In Newtonian cosmology
this important argument has never been formulated explicitly. Indeed, taking
an average value of the Navier-Stokes field operator, the left-hand side of
Eq. (II-10), does not produce the material derivative of an average of the
fluid velocity,
\begin{equation}
\left\langle \frac{\partial u^{i}}{\partial t}+u^{j}\frac{\partial u^{i}}{
\partial x^{j}}\right\rangle \neq \frac{\partial \left\langle
u^{i}\right\rangle }{\partial t}+\left\langle u^{j}\right\rangle \frac{
\partial \left\langle u^{i}\right\rangle }{\partial x^{j}}
\label{navier-stokes-noncomm}
\end{equation}
where $\left\langle \cdot \right\rangle $ stands for an averaging operator.
Thus, similar to the Averaging problem in general relativity one cannot
expect that an arbitrary inhomogeneous Newtonian universe would provide a
solution to the Newtonian averaging problem. A thorough analysis is required
to understand if there exist some locally inhomogeneous Newtonian universes
satisfying the Newtonian cosmological principle on their largest scales.

\section{Averaging and Space-time Measurements}

For an inhomogeneous and/or anisotropic Newtonian universe, the spatial and
temporal dependence of the instant values of the cosmological fluid velocity
$u^{i}=u^{i}(x^{k},t)$ are very complex. Moreover, if the cosmological fluid
flow is turbulent the values of the fluid velocity field are known to be
different each time even if the initial conditions have been set the same
for each fluid configuration. Due to this extreme uncertainty and sharp
changes in space and time of the cosmological fluid velocity field $
u^{i}(x^{k},t)$, the density field $\rho (x^{k},t)$ and the pressure field $
p(x^{k},t)$, it is necessary to use an averaging procedure to introduce more
regular mean values of the fluid characteristic quantities instead of real
fast and sharp varying dynamical fluid fields. These averaged fields may
then be studied analytically by means of usual methods of mathematical
physics.

Another justification for using the averaged quantities for characterization
of the fluid motion comes from the measurement theory. Indeed, every real
act of measurement of the velocity, the density or the pressure of a fluid
particle is carried out by an observer for some interval of time $
T=t_{2}-t_{1}$ during which the particle is moving to another position $
\{x_{2}^{k}\}=\{x_{1}^{k}+\Delta x^{k}\}$. The measurement time interval,
and, correspondingly, the location where a fluid particle moves during the
measurement, depends on the measurement device and conditions. To reduce a
measurement error in the determination of a fluid field below an acceptable
or required value, a measurement may take a considerable time and the
influence of possible random fluctuations in the measured fluid quantity
value is then reduced as necessary. The measured value of the fluid field is
an average over a measurement time interval $T$ and over a space region $
S\subset E^{3}$ which is determined by the precision of the position
resolution of the measurement device. All fluctuations in the quantity under
measurement with the temporal variations much less than the measurement time
and with the spatial variations of much smaller scales than the measurement
device resolution scale are completely suppressed without any trace of them
in the measured average value\footnote{
One should bear in mind, however, that it only concerns the linear
fluctuations of fluid charateristics. Average values of quadratic and
higher-order combinations of fluctuating fields do not vanish in general to
bring the fluid moment, or correlation, functions which are essential for
understanding of the dynamics of moving fluids \cite{Moni-Yagl:1971}, \cite
{Moni-Yagl:1975}, see Section \ref{*stacf}.}. Therefore, a formulation of
Newtonian cosmology in terms of space and time averages of the cosmological
fluid velocity, density and pressure means, from the physical point of view,
that the dynamics of the Newtonian universe is written directly in terms of
observationally relevant variables, and all possible physical effects such
as fluid correlations have been introduced into the theoretical framework
during derivation of such an averaged theory.

\section{Averaging out a Cosmological Fluid Field}

\label{*aocff}

A physical picture of an inhomogeneous Newtonian universe is that any fluid
particle has different values of its instant velocity $u^{i}(x^{j},t)$, the
density $\rho (x^{i},t)$ and the pressure $p(x^{i},t)$ as it moves moving
along its path. All other fluid particles have different values of the
instant velocity, the density and the pressure as they move along their
paths and as each compared with others. In order to describe now a
cosmological fluid configuration in terms of the mean values of fluid
characteristics one should be able to construct the average fields for the
fluid velocity $\left\langle u^{i}(x^{k},t)\right\rangle $, the fluid
density $\left\langle \rho (x^{k},t)\right\rangle $ and the fluid pressure $
\left\langle p(x^{k},t)\right\rangle $. A cosmological fluid configuration
with the average characteristics may have a different dynamics as compared
with the original fluid configuration as far as the average fields are not
equal to the original fluid fields,
\begin{equation}
\left\langle u^{i}(x^{j},t)\right\rangle \neq u^{i}(x^{j},t),\hspace{0.4cm}
\left\langle \rho (x^{i},t)\right\rangle \neq \rho (x^{i},t),\hspace{0.4cm}
\left\langle p(x^{i},t)\right\rangle \neq p(x^{i},t).  \label{macro-micro}
\end{equation}
An equation of motion $x^{i}=x^{i}(\xi ^{j},t)$ for a fluid particle moving
in an inhomogeneous self-gravitating fluid configuration and an equation of
motion $X^{i}=X^{i}(\xi ^{j},t)$ for the same fluid particle moving in an
averaged self-gravitating fluid configuration are expected to be different
because of different solutions of the initial value problems for the the
fluid velocity field $u^{i}(x^{j},t)$, see (I-10) and (II-31),
\begin{equation}
\frac{dx^{i}}{dt}=u^{i}(x^{j},t),\quad x^{i}(0)=\xi ^{i},
\label{velocity-eqs}
\end{equation}
and for the average fluid velocity field $\left\langle
u^{i}(x^{j},t)\right\rangle $ (\ref{macro-micro})
\begin{equation}
\frac{dX^{i}}{dt}=\left\langle u^{i}(x^{j},t)\right\rangle ,\quad
X^{i}(0)=\xi ^{i},  \label{aver-velocity-eqs}
\end{equation}
even if the fluid particle has been given the same initial values $
x^{i}(0)=\xi ^{i}$ and $X^{i}(0)=\xi ^{i}$ in both fluid configurations. So
the paths of the same fluid particle as it is moving within an inhomogeneous
cosmological fluid distribution and within the corresponding averaged
cosmological fluid distribution have to be different in general.

The question of the definition of the average values of classical physical
fields is a delicate one and it has a long history in the classical theory
of fields (for a discussion and references see \cite{Moni-Yagl:1971}, \cite
{Uhle-Ford:1963}, \cite{Lesl:1973} for fluid mechanics, \cite{Lore:1916}-
\cite{Inga-Jami:1985} for the classical macroscopic electrodynamics, and
\cite{Kras:1997}, \cite{Zala:1997}, \cite{Mars-Zala:1997} for general
relativity). In practice, as it has been pointed out above one usually deals
with a space-time average values as resulting from observations and
measurements. Let us consider a quantity $f$ which characterizes the fluid
in the framework of the field description. Generally, for an inhomogeneous
and/or anisotropic cosmological fluid configuration evolving in time such a
quantity $f$ is a function of the Eulerian variables $f=f(x^{i},t)$ with
nonvanishing temporal and spatial derivatives
\begin{equation}
\frac{\partial f}{\partial t}\neq 0,\hspace{0.4cm}\frac{\partial f}{\partial
x^{i}}\neq 0,  \label{f-inhomogeneous}
\end{equation}
which define a temporally and spatially varying field $f(x^{i},t)$ for every
fluid particle. To provide a nontrivial averaged cosmological dynamics the
cosmological fluid field $f(x^{i},t)$ is assumed to satisfy the following
physical hypothesis:\vspace{0.2cm}

\noindent \textbf{The Newtonian cosmological macroscopic hypothesis}\textsc{
\hspace{0.2cm}}\emph{Given a measurement time interval }$T$\emph{\ and a
space region }$S\subset E^{3}$\emph{\ with a volume }$V_{S}$ \emph{
determined by the measurement device resolution, a cosmological fluid field }
$f(x^{i},t)$\emph{\ is supposed to have, at least, either two essentially
different temporal variation scales }$\left( \lambda _{T},L_{T}\right) $
\emph{, or two essentially different spatial variation scales }$\left(
\lambda _{S},L_{S}\right) $\emph{, or both pairs of scales simultaneously,
such as }
\begin{equation}
\lambda _{T}<<T<<L_{T},\hspace{0.4cm}\lambda _{S}<<V_{S}^{-1/3}<<L_{S}
\label{macro-cosmo}
\end{equation}
\emph{where }$\lambda _{T}$\emph{\ and }$\lambda _{S}$\emph{\ are the
temporal and spatial scales of local fluid field fluctuations, and }$L_{T}$
\emph{\ and }$L_{S}$\emph{\ are the temporal and spatial mean cosmological
scales, correspondingly. }\vspace{0.2cm}

Since there is in fact a hierarchy of distinct physical scales in our
Universe \cite{Elli:1984}, the validity of the Newtonian cosmological
macroscopic hypothesis for a fluid field $f(x^{i},t)$ should be checked for
each pair of scales of interest.

Hypotheses analogous to the Newtonian cosmological macroscopic hypothesis
(\ref{macro-cosmo}) are always assumed, very often implicitly, for every
time, space, or space-time averaging scheme (see \cite{Moni-Yagl:1971}, \cite
{Uhle-Ford:1963} for fluid mechanics, \cite{Lore:1916}-\cite{Inga-Jami:1985}
for the classical macroscopic electrodynamics, and \cite{Kras:1997}, \cite
{Zala:1997}, \cite{Mars-Zala:1997} for general relativity). The physical
picture of a Newtonian universe satisfying the Newtonian cosmological
macroscopic hypothesis (\ref{macro-cosmo}) corresponds to our intuitive
understanding of our Universe based on observations and space experiments.
The Universe is seen now as highly inhomogeneous on the scales of different
matter structures which are represented by temporal and/or spatial scales of
local fluid field fluctuations $\lambda _{T}$\textsc{\ and }$\lambda _{S}$.
On its largest scales the Universe shows a definite tendency to homogeneity
and isotropy which are represented by temporal and/or spatial mean
cosmological scales, $L_{T}$\textsc{\ and }$L_{S}$. On some stages of the
evolution of our Universe one can expect, however, that the hypothesis (\ref
{macro-cosmo}) may not be valid or has to be modified. For instance, a
consideration of the Universe evolution at early times when it might have
had a turbulent regime, the cosmological fluid may have more complicated
fluctuation frequency profiles. In this situation, using a mean field
picture and the corresponding averages should thoroughly investigated to
verify whether or not such a consideration is physically relevant and
adequate, see Sections II-6 and \ref{*apnc}.

In dependence of whether the Newtonian cosmological macroscopic hypothesis
(\ref{macro-cosmo}) valid or not for a cosmological fluid distribution, one
can define three possible classes of Newtonian universes.\vspace{0.2cm}

\noindent \textbf{Class LC (Locally inhomogeneous and/or anisotropic
Newtonian universes with the large-scale cosmological dynamics)}\textsc{
\hspace{0.2cm}}\emph{Such Newtonian universes satisfy the Newtonian
cosmological macroscopic hypothesis (\ref{macro-cosmo}) and the averaged
cosmological fluid fields} $\left\langle f(x^{i},t)\right\rangle $ \emph{has
a nontrivial dynamics with either the characteristic temporal variation scale
} $\lambda _{T}$, \emph{or the spatial variation scale} $\lambda _{S}$,
\emph{or with both scales of local fluid field fluctuations with the
condition (\ref{macro-micro}) being held.} \vspace{0.2cm}

This class of Newtonian universes, if it is not empty, would provide a
solution to the Averaging problem, see Section \ref{*apnc}. One of the
Newtonian cosmological models of this class may be a realistic cosmological
model of our Universe in the framework of Newtonian cosmology.\vspace{0.2cm}

\noindent \textbf{Class L (Locally inhomogeneous and/or anisotropic
Newtonian universes without the large-scale cosmological dynamics)}\textsc{
\hspace{0.2cm}}\emph{If a cosmological fluid configuration does not possess
any temporal and/or spatial mean cosmological scales, }$L_{T}$\emph{\ and }$
L_{S}$,\emph{\ when the averages of the cosmological fluid fields vanish and
there is no large-scale cosmological regime of the evolution of such
Newtonian universes, that is,}
\begin{equation}
\left\langle u^{i}(x^{j},t)\right\rangle =0,\hspace{0.4cm}\left\langle \rho
(x^{i},t)\right\rangle =0,\hspace{0.4cm}\left\langle p(x^{i},t)\right\rangle
=0.  \label{macro-micro-micro}
\end{equation}
\vspace{0.2cm}

This class of Newtonian universes has clearly no physical interest when one
is looking for a realistic cosmological model. However, it is of interest in
connection with the Averaging problem since, if it is found, one may be able
to formulate a set of definite constraints on the local matter distribution
necessary for a Newtonian universe to belong to the Class LC.\vspace{0.2cm}

\noindent \textbf{Class C (Newtonian universes with the large-scale
cosmological dynamics only)}\textsc{\hspace{0.2cm}}\emph{If a cosmological
fluid configuration does not possess any temporal and/or spatial scales of
local fluid field fluctuations }$\lambda _{T}$\emph{\ and }$\lambda _{S}$
\emph{, when no nontrivial averaged regime is possible and the average
cosmological fluid fields are equal to the original fluid fields, that is,}
\begin{equation}
\left\langle u^{i}(x^{j},t)\right\rangle =u^{i}(x^{j},t),\hspace{0.4cm}
\left\langle \rho (x^{i},t)\right\rangle =\rho (x^{i},t),\hspace{0.4cm}
\left\langle p(x^{i},t)\right\rangle =p(x^{i},t).  \label{macro-micro-macro}
\end{equation}
\vspace{0.2cm}

This class of Newtonian universes does not seem to be realistic since our
Universe does have temporal and/or spatial scales of local fluid field
fluctuations $\lambda _{T}$\textsc{\ and }$\lambda _{S}$, though a locally
inhomogeneous matter distribution. Again, it is of interest in connection
with the Averaging problem since, if this Class is found, one may be able
formulate a set of definite constraints on the large-scale cosmological
dynamics necessary for a Newtonian universe to belong to the Class LC.

As a fundamental physical consequence of the Newtonian cosmological
macroscopic hypothesis (\ref{macro-cosmo}), an averaging region $S\times T$
of a space-time measurement is considered effectively as a single
\textquotedblleft point\textquotedblright\ for the averaged cosmological
fluid field $\left\langle f(x^{i},t)\right\rangle $. Such regions have been
called \textquotedblleft physically infinitesimally small\textquotedblright\
by Lorentz \cite{Lore:1916} and the results of the measurements must be
insensitive to a choice of a reference point $\left( x^{i},t\right) $ to
which the average $\left\langle f(x^{i},t)\right\rangle $ is prescribed, $
t\in T$, $\{x^{k}\}\in S$. Under the Newtonian cosmological macroscopic
hypothesis (\ref{macro-cosmo}) any measurements of cosmological fluid
quantities over a time interval $T$\ and a space region $S\subset E^{3}$\
with a volume $V_{S}$ determined by the measurement device resolution
guarantee that the errors made while performing such measurements in
different positions $\{x^{i}\}$ within the region $S$, $\{x^{k}\}\in S$, and
different instants of time $t\,$within the interval $T$, $t\in T$, are
always less than $\mathcal{O}(\lambda _{T}/T)$ and $\mathcal{O}(\lambda
_{S}/V_{S}^{-1/3})$ (see \cite{Moni-Yagl:1971} for a discussion and
references therein). The possibility of choosing the averaging time interval
$T$\ and a space region $S\subset E^{3}$\ with a volume $V_{S}$ to be
intermediate between the fluctuating and mean fluid field scales, $\lambda
_{T}$, $\lambda _{S}$ and $L_{T}$, $L_{S}$, assumes that the cosmological
fluid motion may be resolved into a smoothly varying averaged motion with a
very irregular fluctuating motion superimposed on it. There is a
considerable gap between the frequency range characteristic for the mean
motion and fluctuating motion.

In studying the averaged (mean) cosmological fluid configurations there are
three following types of averages of a cosmological fluid field $f(x^{i},t)$
of interest: a time average $\left\langle f(x^{i},t)\right\rangle _{T}$, a
volume space average $\left\langle f(x^{i},t)\right\rangle _{S}$ and an
ensemble, or statistical, average $\left\langle f(x^{i},t)\right\rangle _{E}$
. The first two types of averages can be naturally defined and treated
together as a space-time average $\left\langle f(x^{i},t)\right\rangle _{ST}$
.

\section{The Space-time Averages and Correlations of a Cosmological Fluid}

\label{*stacf}

Given a measurement time interval $T$ and a space region $S\subset E^{3}\ $
determined by the measurement device resolution, a space-time average of the
fluid field $f=f(x^{i},t)$ is defined \cite{Zala:1997}, \cite{Moni-Yagl:1971}
, \cite{Nova:1955},\ \cite{Inga-Jami:1985}, \cite{Mars-Zala:1997} as \vspace{
0.2cm}

\noindent \textbf{The space-time average of a fluid field }$f$\textsc{\ }

\begin{equation}
\left\langle f(x^{i},t)\right\rangle _{ST}=\frac{1}{TV_{S}}
\int_{T}\int_{S}f(x^{i}+x^{\prime i},t+t^{\prime })dt^{\prime }dV^{\prime }
\label{aver-space-time}
\end{equation}
\emph{where $V_{S}$ is 3-volume of the space region $S$,}
\begin{equation}
V_{S}=\int\limits_{S}dV.  \label{aver-space-region}
\end{equation}
\vspace{0.2cm}

\noindent The space-time average (\ref{aver-space-time}) is a well-defined function
\footnote{
Sometimes a continuous weighing function vanishing outside the space-time
region $S\times T$ is used in the definition (\ref{aver-space-time}) where
the integration is then carried out over the whole four-dimensional manifold
defined for the physical confuguration under consideration. Such a weighting
function is completely determined by the properties of a measurement device
and the experimental conditions.} of a reference moment of the absolute time
$t$, $t\in T$, and of a reference position $\{x^{k}\}$, $\{x^{k}\}\in S$,
with the space-time average $\left\langle f(x^{i},t)\right\rangle _{ST}$
being prescribed to $\left( x^{i},t\right) $. If the Newtonian cosmological
macroscopic hypothesis (\ref{macro-cosmo}) is satisfied, the choice of the
reference time and position is arbitrary within the averaging region $
T\times S$. The average value is generally a functional of $T$ and $V_{S}$.

Because the Navier-Stokes-Poisson equations are nonlinear one faces
necessity to deal with averaging products of fluid fields. The space-time
average (\ref{aver-space-time}) for a product of cosmological fluid fields $
f=f(x^{i},t)$\ and $h=h(x^{i},t)$ defines the so-called correlation function.
\vspace{0.2cm}

\noindent \textbf{The two-point second order space-time correlation, or
moment, function of fluid fields }$f$ \textbf{and }$h$\hspace{0.2cm}\emph{
The two-point second order correlation function is the space-time average of
a product of }$f(x^{i},t)$\emph{\ and }$h(y^{j},s)$\textsc{\ }

\begin{equation}
\left\langle f(x^{i},t)h(y^{j},s)\right\rangle _{ST}=\frac{1}{TV_{S}}
\int_{T}\int_{S}f(x^{i}+x^{\prime i},t+t^{\prime })h(y^{j}+x^{\prime
j},s+t^{\prime })dt^{\prime }dV^{\prime }.  \label{aver-space-time-product}
\end{equation}
\vspace{0.2cm}

\noindent The correlation function (\ref{aver-space-time-product}) is a well-defined
function of the reference moments $t$ and $s$ of the absolute time, $t,s\in
T $, and of the reference positions $\{x^{k}\}$ and $\{y^{k}\}$, $
\{x^{k}\},\{y^{k}\}\in S$, with the space-time average $\left\langle
f(x^{i},t)h(y^{j},s)\right\rangle _{ST}$ being prescribed to the pair of
Eulerian variables $\left( x^{i},t\right) $ and $(y^{j},s)$ and symmetric in
these variables
\begin{equation}
\left\langle f(x^{i},t)h(y^{j},s)\right\rangle _{ST}=\left\langle
f(y^{j},s)h(x^{i},t)\right\rangle _{ST}  \label{aver-space-time-product-sym}
\end{equation}

From the physical point of view the two-point space-time average (\ref
{aver-space-time-product}) corresponds to a simultaneous measurement of $f$
and $h$ by an observer with a measurement device with time measurement
interval $T$ and a measurement device resolution region $S\subset E^{3}$. As
far as the Newtonian cosmological macroscopic hypothesis (\ref{macro-cosmo})
is satisfied for the cosmological fluid configuration under consideration a
choice of the reference times $t$ and $s$ and of the reference positions $
\{x^{k}\}$ and $\{y^{k}\}$ is arbitrary within $T$ and $S$. The average
products of fluid fields are known \cite{Moni-Yagl:1971}, \cite{Reyn:1894},
\cite{Kell-Frie:1924}, \cite{Tayl:1921} to be fundamental characteristic
functions responsible for essentially nonlinear phenomenon in evolving
fluids, such as, for example, turbulence and hydrodynamic instability. If
some of such functions do not vanish for a fluid configuration, then the
corresponding fluid fields are said to have correlations.\vspace{0.2cm}

\noindent \textbf{The fluid field correlations}\hspace{0.2cm}\emph{A
cosmological fluid configuration has a central two-point second order
space-time correlation, or moment, function }$C_{ST}^{(2)}(x^{i},t;y^{j},s)$
\emph{, if there are, at least, two cosmological fluid fields }$f(x^{i},t)$
\emph{\ and }$h(x^{i},s)$\emph{\ such that}
\begin{equation}
C_{ST}^{(2)}(x^{i},t;y^{j},s)=\left\langle f(x^{i},t)h(y^{j},s)\right\rangle
_{ST}-\left\langle f(x^{i},t)\right\rangle _{ST}\left\langle
h(y^{j},s)\right\rangle _{ST}\neq 0,  \label{func-correlation}
\end{equation}
\begin{equation}
C_{ST}^{(2)}(x^{i},t;y^{j},s)=C_{ST}^{(2)}(y^{j},s;x^{i},t)
\label{func-correlation-sym}
\end{equation}
\emph{in some open space region }$U\subset E^{3}$\emph{\ for an interval of
time }$\Delta t$, $x^{k},y^{k}\in U$, $t,s\in \Delta t$.\vspace{0.2cm}

One can define the central multi-point higher-order space-time correlation
functions $C_{ST}^{(3)}(x^{i},t;y^{j},s;z^{k},r)$ and so on, whenever it is
necessary for analysis of the dynamics of a Newtonian universe by making
definitions similar to (\ref{aver-space-time-product}) and (\ref
{func-correlation}).

In order to characterize the space-time averaged fluid configuration one
needs to determine now an average field $\left\langle
f(x^{i},t)\right\rangle _{ST}$ and a correlation function field $
\left\langle f(x^{i},t)h(y^{j},s)\right\rangle _{ST}$ for each moments of
time $t$ and $s$ at all positions $\{x^{k}\}$ and $\{y^{k}\}$. From the
physical point of view, that means that it is possible, at least in
principle, to carry out the measurements of the quantities$\ f$ and $h$ for
all instants of time during the evolution of a fluid configuration and at
its each space point. This requires additional assumptions concerning the
measurement interval $T$ and the space region $S$ which are usually made
only tacitly, or they are supposed to be trivial (see a discussion in \cite
{Nova:1955}, \cite{Mars-Zala:1997}). This is, however, a very significant
issue for a correct definition of analytical properties for an average field
(see \cite{Zala:1992}, \cite{Zala:1997}, \cite{Mars-Zala:1997} for a
detailed discussion).\vspace{0.2cm}

\noindent \textbf{The First condition (A covering of the Newtonian
space-time by averaging time intervals and space regions)}\hspace{0.2cm}
\emph{A measurement time interval }$T$\emph{\ and a space region }$S$\emph{\
must be prescribed at every moment of time }$t$\emph{\ and every point point
}$\{x^{k}\}$\emph{\ in order to define an averaged fluid field.}\vspace{0.2cm
}

Given such a covering by the measurement interval $T$ and the space region $
S $, the set of the average values calculated at each moment of time $t$ and
at each position $\{x^{k}\}$ will form average fields $\left\langle
f(x^{i},t)\right\rangle _{ST}$ and $\left\langle
f(x^{i},t)h(y^{j},s)\right\rangle _{ST}$. As noted above, this set
corresponds to the average values of $\ f$ as they are measured for all
instants of time during the evolution of a cosmological fluid configuration
at its each space point.\vspace{0.2cm}

\noindent \textbf{The Second condition (Typical averaging time intervals and
space regions)}\hspace{0.2cm}\emph{All time intervals }$T$\emph{\ and the
space regions }$S$\emph{\ are typical in some defined sense.}\vspace{0.2cm}

They are usually required to be of the same shape and volume, $T=\mathrm{
const}$ and $V_{S}=\mathrm{const}$, such as
\begin{equation}
\frac{\partial T}{\partial t}=0,\quad \frac{\partial T}{\partial x^{i}}
=0,\quad \frac{\partial V_{S}}{\partial t}=0,\quad \frac{\partial V_{S}}{
\partial x^{i}}=0,  \label{covering}
\end{equation}
at each moment of time$\ t$ and at each position $\{x^{k}\}$. For a
Newtonian universe which has the Newtonian space-time geometry, see Sections
II-3 and II-5, one can always satisfy both conditions by placing a
measurement device at all positions $\{x^{k}\}$ in $E^{3}$ at an initial
moment of time $t=t_{0}$ and by the Lie-dragging of the time interval $T$
and the region $S$ along the congruences of the Cartesian coordinate lines $
x^{k}$ and the absolute time $t$ to get a covering\ of the manifold by the
regions $S$ of the same shape and volume and the time intervals of the same
length $T$ at every point of the Newtonian space-time. Another way of
arranging such a covering is to associate a measurement device to all
positions $\{x^{k}\}$ in $E^{3}$ at an initial moment of time $t=t_{0}$ with
fluid particles located at $\{x^{k}\}$ at $t=t_{0}$ and transport the device
to each other position during the evolution of a cosmological fluid
configuration by comoving with the corresponding fluid particles as they
evolve. Both prescriptions must bring the same covering, that is, associate
a standard measurement device with the measurement time interval $T$ and the
measurement resolution space region $S$ with each point $\{x^{k}\}$ at each
instant of time $t$ throughout the cosmological fluid configuration. Three
issues here are of particular importance for notice:

(\emph{A}) Though the second procedure involving transportation along
cosmological fluid paths of moving fluid particles seem to be more physical,
one should bear in mind that this way of construction of a field of the
measurement devices relies on the dynamical evolution of a cosmological
fluid. In many case it does not cause any complications. However, in
situations when the dynamics of an inhomogeneous Newtonian universe exhibits
some specific fluid path configurations, like asymptotic attractors, there
might be some points in the Newtonian space-time where no measurement device
can be brought by this procedure. The first procedure is essentially
kinematic and relies only on the topological and differential structure of
the underlying Newtonian space-time manifold for an inhomogeneous Newtonian
universe. The Eulerian variables $\left( x^{i},t\right) $ are defined
globally on the Newtonian space-time manifold, therefore by the Lie-dragging
of the measurement devices along the coordinate lines $\left( x^{i},t\right)
$ they can be determined for all times $t$ and positions $\{x^{k}\}$. The
Lie-dragging itself is a physically well-defined and meaningful procedure
since it preserves physical scales along the coordinates lines $\left(
x^{i},t\right) $ and does not affect the manifold symmetries and structure.
It does fit perfectly for the purpose of the construction of a covering by
averaging regions without any possible dynamical restrictions.

(\emph{B}) Since time in Newtonian cosmology $t$ is absolute, all clocks of
any observers in an inhomogeneous Newtonian universe are always
synchronized. So arranging the same measurement time intervals $T$ by such a
covering is undoubtedly possible. The situation with a space region $
S\subset E^{3}$\ with a volume $V_{S}$ determined by the measurement device
resolution is more delicate. A typical argument against a possibility to
arrange a covering with space region $S$ of the same shape and volume makes
use of the fact that an evolving cosmological fluid configuration may expand
or shrink during its evolution. However, one can easily show that this
argument does not even appear if all physical hypotheses have been clearly
stated and checked. Indeed, in accordance with the Newtonian cosmological
macroscopic hypothesis (\ref{macro-cosmo}) the characteristic scale $
V_{S}^{-1/3}$ of the space regions $S$ must be always much less than a
characteristic cosmological scale $L_{S}$, $V_{S}^{-1/3}<<L_{S}$. Therefore
any peculiarities of the cosmological fluid evolution of a characteristic
cosmological scale $L_{S}$ cannot affect the measurement device resolution
and change either the shape or the volume of the averaging space regions $S$
placed at different positions at different times. Therefore, such a covering
by the same space regions $S$ is always possible.

(\emph{C}) The values of the space-time averaged fluid field $\left\langle
f(x^{i},t)\right\rangle _{ST}$ calculated for such a covering depend now on $
T$ and $V_{S}$ as free parameters and such an average field becomes
effectively a local single-valued function of the reference time$\ t$ and
position $\{x^{k}\}$ only, and the correlation function $\left\langle
f(x^{i},t)h(y^{j},s)\right\rangle _{ST}$ becomes a local single-valued
function of each pair of variables. This fundamental result and its
consequences for the analytical properties of the space-time averages (\ref
{aver-space-time}) are analyzed in the next Section.

\section{The Reynolds Conditions for Average Fluid Fields}

\label{*rcaff}

In choosing some particular averaging schemes, in addition to the conditions
for the construction of the averages fluid fields, see Section \ref{*stacf},
one must also investigate a possibility to formulate the general
requirements which such a scheme should have. From the point of view of
Newtonian cosmology, the most important of these general requirements is, of
course, that the application of an averaging rule to the
Navier-Stokes-Poisson equations will allow one to get sufficiently simple
equations for the average cosmological fluid fields. Reynolds has found such
a set of conditions while he was using time averages for the derivation of
the averaged Navier-Stokes equations \cite{Reyn:1894}. These conditions have
been reformulated later for any averaging procedure applicable in fluid
mechanics \cite{Moni-Yagl:1971}, \cite{Lesl:1973}, \cite{Stan:1985}, in the
classical macroscopic electrodynamics \cite{Lore:1916}-\cite{Inga-Jami:1985}
and in general relativity \cite{Zala:1992}, \cite{Zala:1993}, \cite
{Mars-Zala:1997}. Let us consider fluid quantities $f=f(x^{i},t)$ and $
h=h(x^{i},t)$ which characterize an inhomogeneous and/or anisotropic
cosmological fluid configuration evolving in time such that both fluid
fields have nonvanishing temporal and spatial derivatives (\ref
{f-inhomogeneous}). The Reynolds conditions are formulated \cite
{Moni-Yagl:1971}, \cite{Lesl:1973}, \cite{Stan:1985} as follows.\vspace{0.2cm
}

\noindent \textbf{The Reynolds Conditions}\textsc{\hspace{0.2cm}}\emph{For
any cosmological fluid fields }$f(x^{i},t)$\emph{\ and }$h(x^{i},t)$\emph{\
the space-time averages }$\left\langle f(x^{i},t)\right\rangle _{ST}$\emph{\
and }$\left\langle h(x^{i},t)\right\rangle _{ST}$\emph{\ must satisfy the
following conditions:}\vspace{0.2cm}

\noindent \emph{(i) the space-time averaging is a linear operation}
\begin{equation}
\left\langle af(x^{i},t)+bh(x^{i},t)\right\rangle _{ST}=a\left\langle
f(x^{i},t)\right\rangle _{ST}+b\left\langle h(x^{i},t)\right\rangle
_{ST},\quad \mathrm{if}\quad a,b=\mathrm{const},  \label{aver-linear}
\end{equation}
\vspace{0.2cm}

\noindent \emph{(ii) the space-time averaging commutes with the partial
differentiation}
\begin{equation}
\frac{\partial }{\partial t}\langle f(x^{i},t)\rangle _{ST}=\langle \frac{
\partial }{\partial t}f(x^{i},t)\rangle _{ST},\quad \frac{\partial }{
\partial x^{i}}\langle f(x^{i},t)\rangle _{ST}=\langle \frac{\partial }{
\partial x^{i}}f(x^{i},t)\rangle _{ST},  \label{aver-commutation}
\end{equation}
\vspace{0.2cm}

\noindent \emph{(iii) the idempotency of the space-time averages}
\begin{align}
\left\langle \langle f(x^{i},t)\rangle _{ST}h(y^{i},s)\right\rangle _{ST}&
=\langle f(x^{i},t)\rangle _{ST}\left\langle h(y^{i},s)\right\rangle
_{ST}\quad \mathrm{or}  \notag \\
\left\langle \langle f(x^{i},t)\rangle _{ST}\right\rangle _{ST}& =\langle
f(x^{i},t)\rangle _{ST}.  \label{aver-idempotent}
\end{align}
\vspace{0.2cm}

One can show that the space-time average (\ref{aver-space-time}) satisfies
the Reynolds conditions, see \cite{Zala:1992}, \cite{Zala:1993}, \cite
{Mars-Zala:1997} for discussion and proofs of relevant issues.

\begin{theorem}[The Reynolds conditions for the space-time averages]
The space-\linebreak time average (\ref{aver-space-time}) is a linear
averaging procedure satisfying the Reynolds condition (\ref{aver-linear}).
If a covering (\ref{covering}) by the averaging time intervals $T$ and the
space regions $S$ is determined through the cosmological fluid
configuration, then

(1*) the space-time average field $\left\langle f(x^{i},t)\right\rangle
_{ST} $ is a local single-valued function of the reference time$\ t$ and the
position $\{x^{k}\}$, which depends on the value of $T$ and $V_{S}$ as
parameters, such as,
\begin{equation}
\left( \frac{\partial }{\partial x^{i}}\frac{\partial }{\partial t}-\frac{
\partial }{\partial t}\frac{\partial }{\partial x^{i}}\right) \left\langle
f(x^{k},t)\right\rangle _{ST}=0\,,\quad \left( \frac{\partial }{\partial
x^{i}}\frac{\partial }{\partial x^{j}}-\frac{\partial }{\partial x^{j}}\frac{
\partial }{\partial x^{i}}\right) \left\langle f(x^{k},t)\right\rangle
_{ST}=0,  \label{aver-locality}
\end{equation}

(2*) the space-time average field $\left\langle f(x^{i},t)\right\rangle
_{ST} $ satisfies the Reynolds condition (\ref{aver-commutation}).
\end{theorem}

\noindent \textbf{Proof.}\hspace{0.4cm}By its definition the space-time
averaging procedure (\ref{aver-space-time}) is a linear operation (\ref
{aver-linear}). Let us now assume that in accordance with the First and
Second conditions of Section \ref{*stacf} a covering (\ref{covering}) has
been arranged for every moment of time$\ t$ at every position $\{x^{i}\}$.
Then one can determine an averaged field $\left\langle
f(x^{i},t)\right\rangle _{ST}$ and define now a spatial partial derivative
of the average field along the coordinate axis $x^{(i)}$ as
\begin{equation}
\frac{\partial }{\partial x^{i}}\left\langle f(x^{k},t)\right\rangle
_{ST}=\lim_{\Delta x^{(i)}\rightarrow 0}\frac{1}{\Delta x^{(i)}}\left[
\left\langle f(x^{k}+\Delta x^{(i)},t)\right\rangle _{ST}-\left\langle
f(x^{k},t)\right\rangle _{ST}\right]   \label{derivative-spatial}
\end{equation}
where the average value $\left\langle f(x^{i}+\Delta x^{(i)},t)\right\rangle
_{ST}$ is taken at the reference point $\{x^{k}+\Delta x^{(k)}\}$ and at the
moment of time$\ t$. Using the definition of the space-time average (\ref
{aver-space-time}), expanding $f(x^{i}+\Delta x^{(i)},t\,)$ into a Taylor
series around the $\{x^{k}\}$ and taking into account the conditions (\ref
{covering}) on the averaging time intervals $T$ and space regions $S$, one
easily derives the commutation formula for the averaging and the spatial
derivative,
\begin{equation}
\frac{\partial }{\partial x^{i}}\langle f(x^{k},t)\rangle _{ST}=\langle
\frac{\partial }{\partial x^{i}}f(x^{k},t)\rangle _{ST},
\label{derivative-spatial2}
\end{equation}
which does not contain explicitly either $T$ or $V_{S}$ as a consequence of
the covering (\ref{covering}). Defining a time derivative of the averaged
field $\left\langle f(x^{i},t\,)\right\rangle _{ST}$ in the same way as (\ref
{derivative-spatial}), one can derive the commutation formula for the
averaging and the time derivative,
\begin{equation}
\frac{\partial }{\partial t}\langle f(x^{k},t)\rangle _{ST}=\langle \frac{
\partial }{\partial t}f(x^{k},t)\rangle _{ST},  \label{derivative-temporal}
\end{equation}
which does not contain explicitly either $T$ or $V_{S}$ as a consequence of
the covering (\ref{covering}). Calculation of the second derivatives of the
averaged field $\left\langle f(x^{i},t)\right\rangle _{ST}$ in accordance
with the definition (\ref{derivative-spatial}) bring the fundamental result
(\ref{aver-locality}). The formula means locality of an average field $
\left\langle f(x^{i},t)\right\rangle _{ST}\,$, that is, it is a local
single-valued function of the reference time$\ t$ and position $\{x^{k}\}$,
which does depend on the value of $T$ and $V_{S}$ as parameters. One can
work with these averages as with usual functions depending on time$\ t$ and
position $\{x^{k}\}$ by differentiating and integrating them, expanding into
series, solving differential and integral equations for them using standard
techniques of mathematical physics. The Reynolds condition (\ref
{aver-commutation}) is given by (\ref{derivative-spatial2}) and (\ref
{derivative-temporal}).\hspace{0.4cm}\textbf{QED}\vspace{0.2cm}

A particular choice of the measurement (averaging) time interval $T$ and the
measurement resolution (averaging) space region $S$ can be fixed at any
convenient stage of analysis. All formalism and the form of the averaged
equations do not depend on the choice of values $T$ and $V_{S}$. When this
choice is made, the average field $\left\langle f(x^{i},t)\right\rangle
_{ST} $ acquires particular values at each point $\{x^{k}\}$ at each instant
of time $t$, which reflects the physical conditions of the measurements and
the cosmological fluid dynamics.

The last Reynolds condition (\ref{aver-idempotent}) is known to be an
approximate condition \cite{Moni-Yagl:1971}, \cite{Zala:1997}, \cite
{Mars-Zala:1997} which holds in Newtonian cosmology due to the Newtonian
cosmological macroscopic hypothesis (\ref{macro-cosmo}).

\begin{corollary}[The idempotency of the space-time averages]
The space-time averages (\ref{aver-space-time}) for a cosmological fluid
configuration satisfying the Newtonian cosmological macroscopic hypothesis
(\ref{macro-cosmo}) are idempotent and the Reynolds condition (\ref
{aver-idempotent}) holds.
\end{corollary}

It should be pointed out here that the physical meaning of the idempotency
(\ref{aver-idempotent}) of the space-time averages (\ref{aver-space-time}) is
that under the Newtonian cosmological macroscopic hypothesis (\ref
{macro-cosmo}) the measurement time interval $T$\ and a space region $
S\subset E^{3}$\ with a volume $V_{S}$ determined by the measurement device
resolution guarantee that the errors made to hold the conditions (\ref
{aver-idempotent}) are always less than the errors of the time and space
measurements, $\mathcal{O}(\lambda _{T}/T)$ and $\mathcal{O}(\lambda
_{S}/V_{S}^{-1/3})$ (see \cite{Moni-Yagl:1971} and references therein).

A satisfactory formal analysis of the idempotency property (\ref
{aver-idempotent}) of the space-time, time and space averages is still
lacking, though it is extensively used for the derivation of the averaged
equations and analysis of various fluid distributions. It has been shown,
however, that there are certain classes of volume averaging procedures which
satisfy all Reynolds conditions (\ref{aver-linear})-(\ref{aver-idempotent}),
but these procedures and corresponding averaging kernels do not have a clear
physical interpretation and they did not find direct applications in fluid
mechanics (see \cite{Kamp:1956}, \cite{Rota:1960} for a discussion and
references).

\section{The Properties and Conditions for the Correlation Functions}

One can now establish the analytic properties of the correlation function $
\left\langle f(x^{i},t)h(y^{j},s)\right\rangle _{ST}$ (\ref
{aver-space-time-product}) and find its properties analogous to the Reynolds
conditions (\ref{aver-linear})-(\ref{aver-idempotent}).

\begin{corollary}[The properties of the space-time correlation functions]
If a covering (\ref{covering}) by the averaging time intervals $T$ and the
space regions $S$ is determined through the cosmological fluid
configuration, then the two-point second order moment function (\ref
{aver-space-time-product}), $\left\langle f(x^{i},t)h(y^{j},s)\right\rangle
_{ST}$, of two cosmological fluid fields $f(x^{i},t)$\ and $h(x^{i},t)$ has
the following properties:

(1*) it is a bilocal single-valued function of the reference times$\ $and
positions, $(x^{k},t)$ and $(y^{k},s)$, which depends on the value of $T$
and $V_{S}$ as parameters, such as all second antisymmetrized derivatives
with respect to all pairs of the variables $(x^{k},t;y^{k},s)$ vanish,

(2*) the two-point second order moment function $\left\langle
f(x^{i},t)h(y^{j},s)\right\rangle _{ST}$ satisfies the conditions of the
partial differentiation,
\begin{align}
\frac{\partial }{\partial x^{k}}\left\langle
f(x^{i},t)h(y^{j},s)\right\rangle _{ST}& =\left\langle \frac{\partial
f(x^{i},t)}{\partial x^{k}}h(y^{j},s)\right\rangle _{ST},  \notag \\
\frac{\partial }{\partial t}\left\langle f(x^{i},t)h(y^{j},s)\right\rangle
_{ST}& =\left\langle \frac{\partial f(x^{i},t)}{\partial t}
h(y^{j},s)\right\rangle _{ST},\quad \mathrm{etc}
\label{aver-commutation-corr}
\end{align}
for all variables $(x^{k},t;y^{k},s)$,

(3*) the two-point second order moment function $\left\langle
f(x^{i},t)h(y^{j},s)\right\rangle _{ST}$ is idempotent
\begin{equation}
\left\langle \left\langle f(x^{i},t)h(y^{j},s)\right\rangle
_{ST}g(z^{k},r)\right\rangle _{ST}=\left\langle
f(x^{i},t)h(y^{j},s)\right\rangle _{ST}\left\langle g(z^{k},r)\right\rangle
_{ST}\quad \mathrm{or}  \label{aver-idempotent-corr0}
\end{equation}
\begin{equation}
\left\langle \left\langle f(x^{i},t)h(y^{j},s)\right\rangle
_{ST}\right\rangle _{ST}=\left\langle f(x^{i},t)h(y^{j},s)\right\rangle
_{ST},  \label{aver-idempotent-corr}
\end{equation}
where $g=g(z^{k},r)$ is another cosmological fluid field.
\end{corollary}

Therefore one can work with the space-time correlation functions as with
usual local functions of many variables depending on reference times$\ $and
positions by differentiating and integrating them, expanding into series,
solving differential and integral equations for them using standard
techniques of mathematical physics.

It should be noted here that the conditions (\ref{aver-commutation-corr})
mean that a partial differentiation of a correlation function produces a
correlation function of the same order including the corresponding partial
derivative of a fluid field.

There is an important asymptotic property \cite{Moni-Yagl:1971} of the
central correlation function $C_{ST}^{(2)}(x^{i},t;y^{j},s)$ (\ref
{func-correlation}), which is of particular interest in Newtonian cosmology.
\vspace{0.2cm}

\noindent \textbf{The asymptotic condition for the central two-point second
order space-time correlation function}\textsc{\hspace{0.2cm}}\emph{For any
two cosmological fluid fields }$f(x^{i},t)$\emph{\ and }$h(x^{i},t)$\emph{\
which are correlated (\ref{func-correlation}) and satisfy the Newtonian
cosmological macroscopic hypothesis (\ref{macro-cosmo}), their central
two-point second order moment function }$C_{ST}^{(2)}(x^{i},t;y^{j},s)$\emph{
\ tends to zero when the distance between the points }$\{x^{i}\}$\emph{\ and
}$\{y^{i}\}$\emph{\ and the time interval between }$t$\emph{\ and }$s$\emph{
\ infinitely grow}\textsc{\ }
\begin{equation}
C_{ST}^{(2)}(x^{i},t;y^{j},s)\rightarrow 0\quad \mathrm{as}\quad \left[
\delta _{ij}(x^{i}-y^{i})(x^{j}-y^{j})\right] ^{1/2}\rightarrow \infty
,\quad \mid s-t\mid \rightarrow \infty .  \label{correlation-asymptotic}
\end{equation}
\vspace{0.2cm}

This property has a very clear physical meaning stating that cosmological
fluid fields are almost independent at extremely remote points in space and
for remote moments of time. So the central correlation functions (\ref
{func-correlation}) for the cosmological fluid fields in the Newtonian
universes of Class LC, see Section \ref{*aocff}, always satisfy the
asymptotic condition (\ref{correlation-asymptotic}). It should be pointed
out, however, that for the Newtonian universes of Classes L and C (\ref
{macro-micro-micro}) and (\ref{macro-micro-macro}) the central correlation
functions (\ref{func-correlation}) may require different asymptotic
conditions. The asymptotic conditions on the correlation functions (\ref
{aver-space-time-product}) should be considered each time separately for
each Newtonian universe.

\section{The Time Averages of a Cosmological Fluid}

\label{*tacf}

Let us consider now the time averages \cite{Moni-Yagl:1971}, \cite{Lesl:1973}
, \cite{Tenn-Luml:1972} which are a particular case of the space-time
average (\ref{aver-space-time}). A time average is defined as a limit of
successive measurements of a fluid field $f=f(x^{i},t)$ over the measurement
time interval $T$ until the fluctuations in its average values become
acceptably small.\vspace{0.2cm}

\noindent \textbf{The time average of a fluid field }$f$
\begin{equation}
\left\langle f(x^{i},t)\right\rangle _{T}=\frac{1}{T}\int
\limits_{t_{1}}^{t_{2}}f(x^{i},t+t^{\prime })dt^{\prime }.  \label{aver-time}
\end{equation}
\vspace{0.2cm}

\noindent Here the time average (\ref{aver-time}) is a well-defined function of a
reference moment of the absolute time $t$ to which the time average $
\left\langle f(x^{i},t)\right\rangle _{T}$ is prescribed, $t\in T$, and the
position $\{x^{i}\}$ where the measurements of $f(x^{i},t)$ are carried out
by an observer with a measurement device resolution region $S$, $
\{x^{i}\}\in S$. The time average (\ref{aver-time}) satisfies the Reynolds
conditions (\ref{aver-linear})-(\ref{aver-idempotent}).

\begin{corollary}[The Reynolds conditions for the time averages]
If a covering (\ref{covering}) by the averaging time intervals $T$ and the
space regions $S$ is determined through the cosmological fluid
configuration, then the time average $\left\langle f(x^{i},t)\right\rangle
_{T}$ is a local single-valued function (\ref{aver-locality}) of the
reference time$\ t$ and the position $\{x^{k}\}$ depending on the value of $
T $ as a parameter and satisfying the Reynolds conditions (\ref{aver-linear}
)-(\ref{aver-idempotent}):

(1*) the time averaging is a linear operation
\begin{equation}
\left\langle af(x^{i},t)+bh(x^{i},t)\right\rangle _{T}=a\left\langle
f(x^{i},t)\right\rangle _{T}+b\left\langle h(x^{i},t)\right\rangle
_{T},\quad \mathrm{if}\quad a,b=\mathrm{const},  \label{aver-linear-time}
\end{equation}

(2*) the time averaging commutes with the partial differentiation
\begin{equation}
\frac{\partial }{\partial t}\langle f(x^{i},t)\rangle _{T}=\langle \frac{
\partial }{\partial t}f(x^{i},t)\rangle _{T},\quad \frac{\partial }{\partial
x^{i}}\langle f(x^{i},t)\rangle _{T}=\langle \frac{\partial }{\partial x^{i}}
f(x^{i},t)\rangle _{T},  \label{aver-commutation-time}
\end{equation}

(3*) the time average is idempotent
\begin{equation}
\left\langle \langle f(x^{i},t)\rangle _{T}h(y^{i},s)\right\rangle
_{T}=\langle f(x^{i},t)\rangle _{T}\left\langle h(y^{i},s)\right\rangle
_{T}\quad \mathrm{or}\quad \left\langle \langle f(x^{i},t)\rangle
_{T}\right\rangle _{T}=\langle f(x^{i},t)\rangle _{T}.
\label{aver-idempotent-time}
\end{equation}
\end{corollary}

Time averages can be applied for the Newtonian universes when a cosmological
fluid distribution satisfies the Newtonian cosmological macroscopic
hypothesis (\ref{macro-cosmo}) for the temporal variation scales $\left(
\lambda _{T},L_{T}\right) $ and there is either a local fluid field spatial
fluctuation scale $\lambda _{S}$, or a characteristic cosmological spatial
scale $L_{S}$. In the first case the cosmological fluid configuration is
completely inhomogeneous in space with a typical spatial variation scale $
\lambda _{S}$ of local fluid field fluctuations, to have the vanishing
space-time average (\ref{aver-space-time}) of a fluid field $f(x^{i},t)$
over a space region $S\subset E^{3}$\ with a volume $V_{S}$ such as $\lambda
_{S}<<V_{S}^{-1/3}$,
\begin{equation}
\left\langle f(x^{i},t)\right\rangle _{ST}=0,\quad \left\langle
f(x^{i},t)\right\rangle _{T}\neq 0.  \label{aver-t-inhom-space}
\end{equation}
In the second case the cosmological fluid configuration is completely
spatially homogeneous (\ref{aver-space-stationary-def}) on the cosmological
spatial scale $L_{S}$. Then the space-time average (\ref{aver-space-time})
of a cosmological fluid field $f(x^{i},t)$ over a space region $S\subset
E^{3}$\ with a volume $V_{S}$ such as $V_{S}^{-1/3}<<L_{S}$ does not change
it (\ref{aver-idempotent}) to smooth $f(x^{i},t)$ effectively only over the
time interval $T$,
\begin{equation}
\left\langle f(x^{i},t)\right\rangle _{ST}=\left\langle f(t)\right\rangle
_{T}.  \label{aver-t-hom-space}
\end{equation}

Let us now consider the condition when the time average for a cosmological
fluid fields $f=f(x^{i},t)$ does not depend on time. The following important
theorem takes place \cite{Moni-Yagl:1971}.

\begin{theorem}[The time average over an infinite time interval]
For any bounded\linebreak\ cosmological fluid quantities $f(x^{i},t)$ and $
h(x^{i},t)$ the time average (\ref{aver-time}) over an infinitely large time
interval $T$ does not depend on time
\begin{equation}
\lim_{T\rightarrow \infty }\left\langle f(x^{i},t)\right\rangle
_{T}=\left\langle f(x^{i})\right\rangle _{T}.  \label{aver-time-stationary}
\end{equation}
The two-point second order time correlation function $\left\langle
f(x^{i},t)h(y^{i},s)\right\rangle _{T}$ defined due to (\ref
{aver-space-time-product}) as
\begin{equation}
\left\langle f(x^{i},t)h(y^{j},s)\right\rangle _{T}=\frac{1}{T}
\int\limits_{t_{1}}^{t_{2}}f(x^{i},t+t^{\prime })h(y^{j},s+t^{\prime
})dt^{\prime }  \label{aver-time-correlation}
\end{equation}
and evaluated over an infinitely large time interval, depends on the
difference $s-t$ only
\begin{equation}
\lim_{T\rightarrow \infty }\left\langle f(x^{i},t)h(y^{j},s)\right\rangle
_{T}=B(x^{i},y^{j};s-t).  \label{aver-time-correlation2}
\end{equation}
\end{theorem}

\noindent \textbf{Proof.}\hspace{0.4cm}The proof is as follows. Let us
consider the same time average over a measurement time interval $T$ at the
reference point $t$, $\left\langle f(x^{i},t)\right\rangle _{T}$, and at
another reference point $s$, $\left\langle f(x^{i},s)\right\rangle _{T}$, $
s,t\in T$, and estimate the difference $\left\langle f(x^{i},s)\right\rangle
_{T}-\left\langle f(x^{i},t)\right\rangle _{T}$. Due to the idempotency (\ref
{aver-idempotent-time}) of the times averages this difference is
approximately zero since under the Newtonian cosmological macroscopic
hypothesis (\ref{macro-cosmo}) the measurement time interval $T$\ guarantee
that the errors made to hold the conditions (\ref{aver-idempotent-time}) are
always less than the errors of the time measurements, $\mathcal{O}(\lambda
_{T}/T)$, see Section \ref{*rcaff}. However, now one would like to find out
a condition when the idempotency holds precisely for the time average
defined as (\ref{aver-time}). Assuming $s>t$ calculation of the difference $
\left\langle f(x^{i},s)\right\rangle _{T}-\left\langle
f(x^{i},t)\right\rangle _{T}$ gives
\begin{equation}
\left\langle f(x^{i},s)\right\rangle _{T}-\left\langle
f(x^{i},t)\right\rangle _{T}=\frac{1}{T}\left[ \int
\limits_{s+t_{1}}^{s+t_{2}}f(x^{i},r)dr-\int
\limits_{t+t_{1}}^{t+t_{2}}f(x^{i},r)dr\right] .  \label{idempotency-rigor}
\end{equation}
For a bounded function of time $f(x^{i},t)$ the difference (\ref
{idempotency-rigor}) becomes infinitely small as $\mid T\mid \rightarrow
\infty $,
\begin{equation}
\left\langle f(x^{i},t)\right\rangle _{T}=\left\langle f(x^{i})\right\rangle
_{T}\quad \mathrm{as}\quad T\rightarrow \infty ,  \label{idempotency-rigor2}
\end{equation}
which proves (\ref{aver-time-stationary}). The proof of (\ref
{aver-time-correlation2}) is straightforward is one considers the limit of
the correlation function $\left\langle f(x^{i},t)h(y^{j},s)\right\rangle
_{T} $ as $T\rightarrow \infty $
\begin{align}
B(x^{i},y^{j};s-t)& =\lim_{T\rightarrow \infty }\left\langle
f(x^{i},t)h(y^{j},s)\right\rangle _{T}=  \notag \\
& \lim_{T\rightarrow \infty }\frac{1}{T}\int
\limits_{t_{1}}^{t_{2}}f(x^{i},t+t^{\prime })h(y^{j},(s-t)+t+t^{\prime
})dt^{\prime }  \label{aver-time-correlation3}
\end{align}
which shows that $\left\langle f(x^{i},t)h(y^{j},s)\right\rangle _{T}$
averaged over an infinitely large time interval can only depend on $(s-t)$,
but not on $s$ and $t$ individually.\hspace{0.4cm}\textbf{QED}\vspace{0.2cm}

The condition (\ref{aver-time-stationary}) can be also easily understood if
one observes that a time average over the whole interval of time allowed by
the evolution of a Newtonian universe must not depend on time. Indeed, an
imaginary measurement of a time dependent physical quantity over the whole
time domain where the quantity is determined gives a number provided such a
time average value is well-defined. The time correlation function (\ref
{aver-time-correlation}) can depend in this case only on the time difference
between the correlated values of the fluid fields (\ref
{aver-time-correlation2}) since the average fields themselves does not
depend on time (\ref{aver-time-stationary}). The higher-order time
correlation functions can be shown all depend only on the time differences,
for example, $B(x^{i},y^{j},z^{k};s-t,r-t)=\lim_{T\rightarrow \infty
}\left\langle f(x^{i},t)h(y^{j},s)g(z^{k},r)\right\rangle _{T}$.

A fluid configuration satisfying everywhere the following conditions for
averaged fluid fields and their time correlation functions over a finite
measurement time interval $T$
\begin{equation}
\left\langle f(x^{i},t)\right\rangle _{T}=\left\langle f(x^{i})\right\rangle
_{T},\quad \left\langle f(x^{i},t)h(y^{j},s)\right\rangle
_{T}=B(x^{i},y^{j};s-t),\quad \mathrm{etc}  \label{aver-time-stationary-def}
\end{equation}
is called stationary.

The physical meaning of stationarity is transparent since all the equations
governing such a fluid motion are time-independent.\vspace{0.2cm}

\noindent \textbf{The Newtonian stationary universes}\textsc{\hspace{0.2cm}}
\emph{A cosmological fluid configuration is stationary if the fluid fields
satisfy the conditions (\ref{aver-time-stationary-def})\ at all positions }$
\{x^{k}\}$\emph{\ during its evolution. \vspace{0.2cm}}

It has been shown in Section II-8 that a particular class of the stationary
Newtonian universes, namely, a static homogeneous and isotropic Newtonian
universe does not exist if the Newtonian cosmological constant vanishes, $
\Lambda =0$.

\section{The Space Averages of a Cosmological Fluid}

\label{*sacf}

Let us consider now the space averages \cite{Moni-Yagl:1971} which are
another particular case of the space-time averages (\ref{aver-space-time}).
A space average is defined as a measurement of a fluid field $f=f(x^{i},t)$
over a space region $S\subset E^{3}$\ with a volume $V_{S}$ determined by
the measurement device resolution such as the fluctuations in its average
value become acceptably small.\emph{\vspace{0.2cm}}

\noindent \textbf{The space average of a fluid field }$f$

\begin{equation}
\left\langle f(x^{i},t)\right\rangle _{S}=\frac{1}{V_{S}}\int_{S}f(x^{i}+x^{
\prime i},t)dV^{\prime }  \label{aver-space}
\end{equation}
\vspace{0.2cm}

\noindent Here the space average (\ref{aver-space}) is a well-defined function of a
reference position $\{x^{k}\}$ to which the space average $\left\langle
f(x^{i},t)\right\rangle _{S}$ is prescribed, $\{x^{k}\}\in S$, and the
moment of time $t$ when the measurements of $f(x^{i},t)$ are carried out by
an observer with a measurement time interval $T$, $t\in T$. The space
average (\ref{aver-space}) satisfies the Reynolds conditions (\ref
{aver-linear})-(\ref{aver-idempotent}).

\begin{corollary}[The Reynolds conditions for the space averages]
If a covering (\ref{covering}) by the averaging time intervals $T$ and the
space regions $S$ is determined, then the space average $\left\langle
f(x^{i},t)\right\rangle _{S}$ is a local single-valued function (\ref
{aver-locality}) of the reference time$\ t$ and position $\{x^{k}\}$
depending on the value of $V_{S}$ as a parameter and satisfying the Reynolds
conditions (\ref{aver-linear})-(\ref{aver-idempotent}):

(1*) the space averaging is a linear operation
\begin{equation}
\left\langle af(x^{i},t)+bh(x^{i},t)\right\rangle _{S}=a\left\langle
f(x^{i},t)\right\rangle _{S}+b\left\langle h(x^{i},t)\right\rangle
_{S},\quad \mathrm{if}\quad a,b=\mathrm{const},  \label{aver-linear-space}
\end{equation}

(2*) the space averaging commutes with the partial differentiation
\begin{equation}
\frac{\partial }{\partial t}\langle f(x^{i},t)\rangle _{S}=\langle \frac{
\partial }{\partial t}f(x^{i},t)\rangle _{S},\quad \frac{\partial }{\partial
x^{i}}\langle f(x^{i},t)\rangle _{S}=\langle \frac{\partial }{\partial x^{i}}
f(x^{i},t)\rangle _{S},  \label{aver-commutation-space}
\end{equation}

(3*) the space average is idempotent
\begin{equation}
\left\langle \langle f(x^{i},t)\rangle _{S}h(y^{i},s)\right\rangle
_{S}=\langle f(x^{i},t)\rangle _{S}\left\langle h(y^{i},s)\right\rangle
_{S}\quad \mathrm{or}\quad \left\langle \langle f(x^{i},t)\rangle
_{S}\right\rangle _{S}=\langle f(x^{i},t)\rangle _{S}.
\label{aver-idempotent-space}
\end{equation}
\end{corollary}

Space averages can be applied for the Newtonian universes when a
cosmological fluid distribution satisfies the Newtonian cosmological
macroscopic hypothesis (\ref{macro-cosmo}) for the spatial variation scales $
\left( \lambda _{S},L_{S}\right) $ and there is either a local fluid field
temporal fluctuation scale $\lambda _{T}$, or a characteristic cosmological
temporal scale $L_{T}$. In the first case the cosmological fluid
configuration is completely inhomogeneous in time with a typical temporal
variation scale $\lambda _{T}$ of local fluid field fluctuations to have the
vanishing space-time average (\ref{aver-space-time}) of a fluid field $
f(x^{i},t)$ with a time measurement interval $T$ such as $\lambda _{T}<<T$
\begin{equation}
\left\langle f(x^{i},t)\right\rangle _{ST}=0,\quad \left\langle
f(x^{i},t)\right\rangle _{S}\neq 0.  \label{aver-s-t-inhom-space}
\end{equation}
In the second case the cosmological fluid configuration is stationary (\ref
{aver-time-stationary-def}) on the cosmological spatial scale $L_{T}$. Then
the space-time average (\ref{aver-space-time}) of a cosmological fluid field
$f(x^{i},t)$ over a time measurement interval $T$ such as $T<<L_{S}$ does
not change it (\ref{aver-idempotent}) to smooth $f(x^{i},t)$ effectively
only over a space region $S$
\begin{equation}
\left\langle f(x^{i},t)\right\rangle _{ST}=\left\langle
f(x^{i})\right\rangle _{S}.  \label{aver-s-t-hom-space}
\end{equation}

Let us now consider the condition when the space average for a cosmological
fluid field $f=f(x^{i},t)$ does not depend on the reference point $\{x^{i}\}$
. The following important theorem takes place \cite{Moni-Yagl:1971}.

\begin{theorem}[The space averages over the whole space]
For any bounded cosmological fluid quantities $f(x^{i},t)$ and $h(x^{i},t)$
the space average (\ref{aver-space}) over the whole space $S=E^{3}$ does not
depend on a position $\{x^{i}\}$
\begin{equation}
\lim_{V_{S}\rightarrow \infty }\left\langle f(x^{i},t)\right\rangle
_{S}=\left\langle f(t)\right\rangle _{S}.  \label{aver-space-stationary}
\end{equation}
The two-point second order space correlation function $\left\langle
f(x^{i},t)h(y^{i},s)\right\rangle _{S}$ defined due to (\ref
{aver-space-time-product}) as
\begin{equation}
\left\langle f(x^{i},t)h(y^{j},s)\right\rangle _{S}=\frac{1}{V_{S}}
\int_{S}f(x^{i}+x^{\prime i},t)h(y^{j}+x^{\prime i},s)dV^{\prime }
\label{aver-space-correlation}
\end{equation}
and evaluated over the whole space $S=E^{3}$ depends on the difference $
y^{j}-x^{i}$ only
\begin{equation}
\lim_{V_{S}\rightarrow \infty }\left\langle
f(x^{i},t)h(y^{j},s)\right\rangle _{S}=B(y^{j}-x^{i};t,s).
\label{aver-space-correlation2}
\end{equation}
\end{theorem}

\noindent \textbf{Proof.}\hspace{0.4cm}The proof can be easily accomplished
in a completely analogous manner as the proof of the similar properties for
the time averages (\ref{aver-time-stationary}) and (\ref
{aver-time-correlation2}).\hspace{0.4cm}\textbf{QED}\vspace{0.2cm}

The condition (\ref{aver-space-stationary}) can be also easily understood if
one observes that a space average over the whole space $E^{3}$ allowed by
the definition of a Newtonian universe must not depend on spatial
coordinates. Indeed, an imaginary measurement of a position dependent
physical quantity over the whole space where the quantity is determined
gives a number provided such a space average value is well-defined. The
space correlation function (\ref{aver-space-correlation}) can depend in this
case only on a difference between the correlated values of the fluid fields
(\ref{aver-space-correlation2}) since the average fields themselves do not
depend on a space position (\ref{aver-space-stationary}). The higher-order
space correlation functions can be shown all depend only on the time
differences, for example, $B(y^{i}-x^{i},z^{j}-x^{j};t,s,r)=\lim_{V_{S}
\rightarrow \infty }\left\langle f(x^{i},t)h(y^{j},s)g(z^{k},r)\right\rangle
_{S}$.

A fluid configuration satisfying everywhere the conditions
\begin{equation}
\left\langle f(x^{i},t)\right\rangle _{S}=\left\langle f(t)\right\rangle
_{S},\quad \left\langle f(x^{i},t)h(y^{j},s)\right\rangle
_{S}=B(y^{i}-x^{i};t,s),\quad \mathrm{etc}  \label{aver-space-stationary-def}
\end{equation}
for the average fluid characteristic fields and their space correlation
functions over a space measurement device resolution region $S\subset E^{3}$
\ is called homogeneous. The physical meaning of homogeneity is transparent
since all the equations governing such a fluid motion are independent of
fluid positions $\{x^{i}\}$.\vspace{0.2cm}

\noindent \textbf{The Newtonian homogeneous universes}\textsc{\hspace{0.2cm}}
\emph{A cosmological fluid configuration is homogeneous if the fluid fields
satisfy the conditions (\ref{aver-space-stationary-def})\ at all positions }$
\{x^{k}\}$ \emph{during its evolution. }\vspace{0.2cm}

It has been shown in Section II-8 that a particular class of homogeneous
Newtonian universes, namely, a static homogeneous and isotropic Newtonian
universe does not exist if the cosmological constant vanishes, $\Lambda =0$.
The class of homogeneous and isotropic Newtonian cosmologies has been
studied in Sections II-6 and II-12.

It is important to compare the formal definition (\ref
{aver-space-stationary-def}) with the formulation of the Newtonian
cosmological principle, see Section II-6, which defines the homogeneity
condition for a cosmological fluid configuration in terms of the data
measured by a privileged family of observers. The homogeneity conditions
(\ref{aver-space-stationary-def}) set precise constraints on the character of
the averaged fluid fields as they are measured by the observers endowed with
measurement devices having the same measurement time interval $T$ and the
same measurement device resolution determined by a space region $S$ (\ref
{covering}).

\section{The Space-time Averaging out of Material Derivatives}

\label{*stamd}

In derivation of the averaged systems of the Navier-Stokes-Poisson equations
(II-4), (II-8), (II-9) and (II-10), or the Navier-Stokes-Poisson equations
in terms of kinematic quantities (II-4), (II-8), (II-12), (II-51)-(II-56),
(II-59)-(II-61), (I-64)-(II-66), (II-69), (II-28) and (II-70) one needs to
consider carefully the problem of taking an average value of the
Navier-Stokes field operator, that is, the left-hand side of Eq. (II-10). As
it has been pointed out in the Section \ref{*apnc} an averaging out of this
operator (\ref{navier-stokes-noncomm}) does not give the material derivative
of an averaged fluid velocity. Let us consider now the space-time averages $
\left\langle u^{i}(x^{k},t)\right\rangle _{ST}$ (\ref{aver-space-time}) of
the cosmological fluid velocity\textsc{\ }$u^{i}(x^{k},t)$ and make use of
the Reynolds conditions (\ref{aver-linear})-(\ref{aver-idempotent}) after
taking the space-time average of the Navier-Stokes field operator, that is,
the material derivative of the fluid velocity. One arrives at the expression
\begin{gather}
\frac{\partial \left\langle u^{i}(x^{k},t)\right\rangle _{ST}}{\partial t}
+\left\langle u^{j}(x^{k},t)\frac{\partial u^{i}(x^{k},t)}{\partial x^{j}}
\right\rangle _{ST}\neq  \notag \\
\frac{\partial \left\langle u^{i}(x^{k},t)\right\rangle _{ST}}{\partial t}
+\left\langle u^{j}(x^{k},t)\right\rangle _{ST}\frac{\partial \left\langle
u^{i}(x^{k},t)\right\rangle _{ST}}{\partial x^{j}}
\label{navier-stokes-noncomm2}
\end{gather}
which again is not equal to the material derivative of the averaged
velocity $\left\langle u^{i}(x^{k},t)\right\rangle _{ST}$ because of the
presence of the single-point second order space-time correlation function of
the fluid velocity and its spatial derivative $\left\langle
u^{j}(x^{k},t)u_{,j}^{i}(x^{k},t)\right\rangle _{ST}$. Therefore, a formula
for performing averaging out the Navier-Stokes field operator is definitely
very important as it will affect the structure of the field operator of the
averaged Navier-Stokes equation and finally that of the averaged
Navier-Stokes-Poisson equations.

In an inhomogeneous Newtonian universe any fluid particle has different
values of its instant velocity $u^{i}(x^{j},t)$, the density $\rho (x^{i},t)$
and the pressure $p(x^{i},t)$ as it moves along its path. A cosmological
fluid configuration with the space-time average fields for the velocity $
\left\langle u^{i}(x^{j},t)\right\rangle _{ST}$, the density $\left\langle
\rho (x^{i},t)\right\rangle _{ST}$ and the pressure $\left\langle
p(x^{i},t)\right\rangle _{ST}$ may have a different dynamics as compared
with the original fluid configuration since the average fields are not equal
to the original fluid fields (\ref{macro-micro}) in general, see Section \ref
{*aocff}. Indeed, a fluid particle moving in an inhomogeneous
self-gravitating fluid configuration with the velocity $u^{i}(x^{i},t)$ has
an equation of motion (I-1),
\begin{equation}
x^{i}=x^{i}(\xi ^{j},t),  \label{flow}
\end{equation}
as a solution to the initial value problem (\ref{velocity-eqs}). Due to the
Newtonian cosmological macroscopic hypothesis (\ref{macro-cosmo}), an
averaging region $S\times T$ is considered effectively as a single
\textquotedblleft point\textquotedblright\ for the averaged cosmological
fluid field $\left\langle f(x^{i},t)\right\rangle _{ST}$, see Section \ref
{*aocff}. Therefore, one can consider now the same fluid particle as moving
in the averaged self-gravitating fluid cosmological configuration with the
average velocity $\left\langle u^{i}(x^{j},t)\right\rangle _{ST}$ such as
(\ref{macro-micro})
\begin{equation}
\left\langle u^{i}(x^{j},t)\right\rangle _{ST}\neq u^{i}(x^{j},t).
\label{velocity-aver}
\end{equation}
The equation of motion
\begin{equation}
X^{i}=X^{i}(\xi ^{j},t)  \label{flow-aver}
\end{equation}
of the fluid particle moving in the averaged cosmological fluid
configuration is different from the equation of motion (\ref{flow}) for the
fluid particle moving in an inhomogeneous cosmological configuration with
the velocity $u^{i}(x^{k},t)$ because a solution to the initial value
problem (\ref{aver-velocity-eqs}),
\begin{equation}
\frac{dX^{i}}{dt}=\left\langle u^{i}(x^{j},t)\right\rangle _{ST},\quad
X^{i}(0)=\xi ^{i},  \label{velocity-eqs-aver}
\end{equation}
is different from that to (\ref{velocity-eqs}) for the same initial values $
x^{i}(0)=\xi ^{i}$ and $X^{i}(0)=\xi ^{i}$. So the paths of the same fluid
particle as it is considered moving within an inhomogeneous cosmological
fluid distribution and within the corresponding averaged cosmological fluid
distribution have to be different in general.

A fluid quantity $f$ in the framework of the field description is a function
of the Eulerian variables, $f=f(x^{i},t)$, and it is also a function of the
material variables, $f=f(\xi ^{i},t)$, such as
\begin{equation}
f=f[x^{i}(\xi ^{j},t),t]\hspace{0.4cm}\mathrm{or}\hspace{0.4cm}f=f[\xi
^{i}(x^{j},t),t],  \label{field}
\end{equation}
where functions
\begin{equation}
\xi ^{i}=\xi ^{i}(x^{j},t)  \label{inverse-flow}
\end{equation}
define the initial position $\{\xi ^{i}\}$, $\xi ^{i}=x^{i}(0)$, of the
fluid particle which is at any position $\{x^{i}\}$ at a moment of time $t$,
see Section I-5. The average fluid field $\left\langle f\right\rangle
=\left\langle f(x^{i},t)\right\rangle _{ST}$ is now a function of the
Eulerian variables (\ref{flow-aver}) and the material variables
\begin{equation}
\xi ^{i}=\xi ^{i}(X^{j},t),  \label{inverse-flow-aver}
\end{equation}
such as
\begin{equation}
\left\langle f\right\rangle =\left\langle f\right\rangle [X^{i}(\xi
^{j},t),t]\hspace{0.4cm}\mathrm{or}\hspace{0.4cm}\left\langle f\right\rangle
=\left\langle f\right\rangle [\xi ^{i}(X^{j},t),t],  \label{field-aver}
\end{equation}
the functions (\ref{field-aver}) being related by Eqs. (\ref{flow-aver}) and
(\ref{inverse-flow-aver}) considered as the laws of change of variables.
Then the change in the quantity $\left\langle f\right\rangle $ in course of
the averaged fluid motion can be characterized by the two different time
derivatives
\begin{equation}
\frac{\partial \left\langle f\right\rangle }{\partial t}\equiv \frac{
\partial \left\langle f\right\rangle (X^{i},t)}{\partial t}_{\mid X^{i}=
\mathrm{const}}\hspace{0.4cm}\mathrm{and}\hspace{0.4cm}\frac{d\left\langle
f\right\rangle }{dt}\equiv \frac{\partial \left\langle f\right\rangle (\xi
^{i},t)}{\partial t}_{\mid \xi ^{i}=\mathrm{const}}.  \label{time-drs-aver}
\end{equation}
The partial derivative $\partial \left\langle f\right\rangle /\partial t$
gives the rate of change of $\left\langle f\right\rangle $ with respect to a
fixed position $\{X^{i}\}$, while the material, or convective, derivative $
d\left\langle f\right\rangle /dt$ measures the rate of change of $
\left\langle f\right\rangle $ with respect to a moving fluid particle.

The material derivative of the position of a fluid particle defined by the
second Eq. (\ref{time-drs-aver}) for $\left\langle f\right\rangle =X^{i}$ is
called its average fluid velocity $\left\langle u^{i}\right\rangle $
\begin{equation}
\left\langle u^{i}\right\rangle (\xi ^{j},t)=\frac{dX^{i}}{dt}\equiv \frac{
\partial X^{i}(\xi ^{j},t)}{\partial t}_{\mid \xi ^{i}=\mathrm{const}},
\hspace{0.4cm}\left\langle u^{i}\right\rangle (\xi ^{j},t)=\left\langle
u^{i}\right\rangle (X^{j},t)=\left\langle u^{i}(x^{j},t)\right\rangle _{ST}.
\label{aver-velocity}
\end{equation}
The notion of the average fluid velocity $\left\langle u^{i}\right\rangle $
(\ref{aver-velocity}) plays a fundamental role in the field formulation of
the fluid mechanics of the averaged flows in the same way as the fluid
velocity $u^{i}$ plays a fundamental role in the field formulation of fluid
mechanics of the inhomogeneous flows, see Section II-5.

With the definitions of time derivatives (\ref{time-drs-aver}) one can
calculate the material derivative of any averaged quantity $\left\langle
f\right\rangle $ as
\begin{align}
\frac{d\left\langle f\right\rangle }{dt}& =\left[ \frac{\partial }{\partial t
}\left\langle f\right\rangle (\xi ^{i},t)\right] _{\mid \xi ^{i}=\mathrm{
const}}=\left\{ \frac{\partial }{\partial t}\left\langle f\right\rangle
[X^{i}(\xi ^{j},t),t]\right\} _{\mid \xi ^{i}=\mathrm{const}}=  \notag \\
& \frac{\partial \left\langle f\right\rangle }{\partial t}_{\mid X^{i}=
\mathrm{const}}+\frac{\partial \left\langle f\right\rangle }{\partial X^{i}}
\left[ \frac{\partial X^{i}(\xi ^{j},t)}{\partial t}\right] _{\mid \xi ^{i}=
\mathrm{const}}  \label{material-dr1-aver}
\end{align}
that can be written in the following form by using the average fluid
velocity defined by Eq. (\ref{aver-velocity}):
\begin{equation}
\frac{d\left\langle f\right\rangle }{dt}=\frac{\partial \left\langle
f\right\rangle }{\partial t}+\left\langle u^{i}\right\rangle \frac{\partial
\left\langle f\right\rangle }{\partial X^{i}}\hspace{0.4cm}\mathrm{or}
\hspace{0.4cm}\frac{d\left\langle f\right\rangle }{dt}=\frac{\partial
\left\langle f\right\rangle }{\partial t}+\left\langle u^{i}\right\rangle
\frac{\partial \left\langle f\right\rangle }{\partial x^{i}}.
\label{material-dr2-aver}
\end{equation}
This formula relates the material and spatial derivatives through the
average velocity field in the field picture of the average fluid motion and
it expresses the rate of change in $\left\langle f\right\rangle $ with
respect to a moving fluid particle located at a position $\{X^{i}\}$ at time
$t$. It is completely analogous to the material derivative of the fluid
field $f(x^{i},t)$ (I-12), see Section I-6,
\begin{equation}
\frac{df}{dt}=\frac{\partial f}{\partial t}+u^{i}\frac{\partial f}{\partial
x^{i}},  \label{material-dr2}
\end{equation}
where it relates the material and spatial derivatives through the velocity
field $u^{i}(x^{j},t)$ in the field description of fluid motion, see Section
I-5, and it expresses the rate of change in $f(x^{j},t)$ with respect to a
moving fluid particle located at a position $\{x^{i}\}$ at time $t$.

The Eulerian coordinates $(X^{i},t)$ of the averaged cosmological fluid
configuration correspond to the Eulerian coordinates $(x^{i},t)$ of an
inhomogeneous cosmological fluid configuration since the averaged value $
\left\langle f(x^{i},t)\right\rangle _{ST}$ has been prescribed to a
reference time $t$ and a position $\{x^{i}\}$ which are the same moment of
time $t$ and the same position $\{X^{i}\}=\{x^{i}\}$. One must distinguish
between the equations of motion of fluid particles moving in an
inhomogeneous fluid distribution and in the corresponding average fluid
distribution averaging out particular relations, but one can always use the
spatial coordinates $\{x^{i}\}$ instead of $\{X^{i}\}$ in the final averaged
formulae (\ref{material-dr2-aver}).

One can now prove the following important theorem establishing the rule for
space-time averaging of the material derivatives.

\begin{theorem}[The space-time average of the material derivative]
If a covering\linebreak\ (\ref{covering}) by the averaging time intervals $T$
and the space regions $S$ is determined through the cosmological fluid
configuration, then the space-time average (\ref{aver-space-time}) of the
material derivative of a cosmological fluid field $f(x^{i},t)$ is given by
the formula
\begin{equation}
\left\langle \frac{df}{dt}\right\rangle _{ST}=\left\langle \frac{\partial f}{
\partial t}+u^{i}\frac{\partial f}{\partial x^{i}}\right\rangle _{ST}=\frac{
d\left\langle f\right\rangle }{dt}+\left\langle u^{i}\frac{\partial f}{
\partial x^{i}}\right\rangle -\left\langle u^{i}\right\rangle \frac{\partial
\left\langle f\right\rangle }{\partial x^{i}}  \label{aver-material}
\end{equation}
where
\begin{equation}
\left\langle u^{i}\frac{\partial f}{\partial x^{i}}\right\rangle
=\left\langle u^{i}\frac{\partial f}{\partial x^{i}}\right\rangle
(x^{j},t)=\left\langle u^{i}(x^{j},t)\frac{\partial f(x^{k},t)}{\partial
x^{i}}\right\rangle _{ST}  \label{material-correlation}
\end{equation}
is the single-point second order space-time correlation function of the
fluid velocity $u^{i}(x^{j},t)$ and its spatial derivative $\partial
f(x^{i},t)/\partial x^{i}$.
\end{theorem}

\noindent \textbf{Proof.}\hspace{0.4cm}The proof is straightforward by
taking the space-time average (\ref{aver-space-time}) of the material
derivative of a cosmological fluid field $f=f(x^{i},t)$ with using the
Reynolds conditions (\ref{aver-commutation}), the definitions of the average
cosmological fluid field $\left\langle f\right\rangle $ (\ref{field-aver})
and the average cosmological fluid velocity $\left\langle u^{i}\right\rangle
$, the definition and expression for the material derivative of the average
fluid field (\ref{aver-velocity}) and (\ref{material-dr2-aver}) and the
introduction of the single-point second order space-time correlation
function (\ref{material-correlation}).\hspace{0.4cm}\textbf{QED}\vspace{0.2cm
}

There are two equivalent useful forms of the formula (\ref{aver-material}):
\begin{equation}
\left\langle \frac{df}{dt}\right\rangle _{ST}=\frac{\partial \left\langle
f\right\rangle }{\partial t}+\left\langle u^{i}\frac{\partial f}{\partial
x^{i}}\right\rangle ,  \label{aver-material2}
\end{equation}
\begin{equation}
\left\langle \frac{df}{dt}\right\rangle _{ST}=\frac{d\left\langle
f\right\rangle }{dt}+C^{(2)}\left[ u^{i},\frac{\partial f}{\partial x^{i}}
\right] ,\hspace{0.4cm}C^{(2)}\left[ u^{i},\frac{\partial f}{\partial x^{i}}
\right] =\left\langle u^{i}\frac{\partial f}{\partial x^{i}}\right\rangle
-\left\langle u^{i}\right\rangle \frac{\partial \left\langle f\right\rangle
}{\partial x^{i}},  \label{aver-material3}
\end{equation}
where $C^{(2)}\left[ u^{i},f_{,i}\right] $ is the central single-point
second order moment function of the fluid velocity $u^{i}(x^{j},t)$ and the
spatial derivative $\partial f(x^{i},t)/\partial x^{j}$ of the fluid field $
f(x^{i},t)$.

The formulae (\ref{aver-material})-(\ref{aver-material3}) are of fundamental
significance in the derivation of the averaged Navier-Stokes-Poisson
equations. An immediate result is their application to the space-time
averaging out of the field operator of the Navier-Stokes equation (\ref
{navier-stokes}).

\begin{corollary}[The averaged Navier-Stokes field operator]
If a covering (\ref{covering}) by the averaging time intervals $T$ and the
space regions $S$ is determined through the cosmological fluid
configuration, then the space-time average (\ref{aver-space-time}) of the
field operator of the Navier-Stokes equation (II-10) has the following form:
\begin{equation}
\left\langle \frac{du^{i}}{dt}\right\rangle _{ST}=\left\langle \frac{
\partial u^{i}}{\partial t}+u^{k}\frac{\partial u^{i}}{\partial x^{k}}
\right\rangle _{ST}=\frac{d\left\langle u^{i}\right\rangle }{dt}
+\left\langle u^{k}\frac{\partial u^{i}}{\partial x^{k}}\right\rangle
-\left\langle u^{k}\right\rangle \frac{\partial \left\langle
u^{i}\right\rangle }{\partial x^{k}}  \label{navier-stokes-aver}
\end{equation}
where
\begin{equation}
\left\langle u^{j}\frac{\partial u^{i}}{\partial x^{j}}\right\rangle
=\left\langle u^{j}\frac{\partial u^{i}}{\partial x^{j}}\right\rangle
(x^{k},t)=\left\langle u^{j}(x^{k},t)\frac{\partial u^{i}(x^{k},t)}{\partial
x^{j}}\right\rangle _{ST}  \label{navier-stokes-aver-corr}
\end{equation}
is the single-point second order moment function of the fluid velocity $
u^{i}(x^{j},t)$ and its spatial derivative $\partial u^{i}(x^{k},t)/\partial
x^{j}$.
\end{corollary}

Thus, the averaged Navier-Stokes field operator has changed its structure
under the space-time averaging (\ref{aver-space-time}) as it has been
suggested by a heuristic argument (\ref{navier-stokes-noncomm2}). The
averaging rule (\ref{navier-stokes-aver}) plays a central role in
establishing the form of the averaged Navier-Stokes-Poisson equations.

\section{Averaged Material Derivatives and the Reynolds Transport Theorem}

The derivation of the formulae (\ref{aver-material})-(\ref{aver-material3})
for a space-time averaging of the material derivatives of cosmological fluid
fields has required a thorough analysis of the averaged fluid motion, see
Section \ref{*stamd}. The foundation of these formulae lies in the
possibility to arrange the covering (\ref{covering}). Then in accordance
with the First and Second conditions, see Section \ref{*stacf}, for a
covering of Newtonian space-time by the averaging time intervals $T$ and the
space regions $S$ typical in some defined sense, one can determine the
space-time average field $\left\langle f(x^{i},t)\right\rangle _{ST}$ which
is a local single-valued function of the reference time$\ t$ and position $
\{x^{k}\}$, which depends on the value of $T$ and $V_{S}$ as parameters and
satisfies the Reynolds conditions (\ref{aver-linear})-(\ref{aver-idempotent}
), see Sections \ref{*stacf} and \ref{*rcaff}. The covering (\ref{covering})
has been shown to be consistent with the Newtonian cosmological macroscopic
hypothesis (\ref{macro-cosmo}) and it associates a standard measurement
device with the measurement time interval $T$ and the measurement resolution
space region $S$ with each point $\{x^{k}\}$ at each instant of time $t$.
This is the physical foundation of the Reynolds conditions which are always
assumed for derivation of the averaged equations and relations for
space-time, time, space and ensemble averages in fluid mechanics \cite
{Moni-Yagl:1971}, \cite{Lesl:1973}, \cite{Stan:1985}, the Maxwell
macroscopic electrodynamics \cite{Lore:1916}-\cite{Inga-Jami:1985} and
general relativity \cite{Zala:1992}, \cite{Zala:1993}, \cite{Mars-Zala:1997}.

Let us now consider a covering of Newtonian space-time by the measurement
time intervals $T$ and the space regions $S\subset E^{3}\ $determined by the
measurement device resolution, which are taken to be typical in different
than the covering (\ref{covering}) sense. If one requires a more general
covering
\begin{equation}
\frac{\partial T}{\partial t}=0,\quad \frac{\partial T}{\partial x^{i}}
=0,\quad \frac{\partial V_{S}}{\partial t}\neq 0,\quad \frac{\partial V_{S}}{
\partial x^{i}}\neq 0  \label{covering2}
\end{equation}
at each moment of time$\ t$ and position $\{x^{k}\}$ that means $T=\mathrm{
const}$ and $V_{S}\neq \mathrm{const}$. In fact, it is actually impossible
to require a covering other than $T=\mathrm{const}$ with the two first
conditions (\ref{covering2}) since the time is absolute in Newtonian
cosmology. A choice of space regions $S\subset E^{3}\ $such as $V_{S}\neq
\mathrm{const}$ with the two last conditions (\ref{covering2}) held is
possible formally. Such a choice associates a measurement device with its
resolution space region $S$ at each point $\{x^{k}\}$ at each instant of
time $t$ with its characteristic scale $V_{S}^{-1/3}$ changing in dependence
of a position in space. In order to have $V_{S}\neq \mathrm{const}$, the
Newtonian cosmological macroscopic hypothesis (\ref{macro-cosmo}) should be
violated by two possible ways, namely, either $\lambda _{S}\simeq
V_{S}^{-1/3}<<L_{S}$, or $\lambda _{S}<<V_{S}^{-1/3}\simeq L_{S}$. In the
first case the change in the space regions $S$ is due to a spatial scale $
\lambda _{S}$ of the local cosmological fluid field fluctuations. It cannot
be taken seriously from the physical point of view since observers would be
affected by the peculiarities of the fast and sharply changing local fields
and any space averaging cannot be performed. In the second case the change
in the space regions $S$ is due to a spatial cosmological scale $L_{S}$ and
the observers can perform space measurements in principle. However, the
measurement space resolution scale $V_{S}^{-1/3}$ is of the same order as a
spatial mean cosmological scale that implies, as a matter of fact, that the
space averaging does not resolve any cosmological scale at all. Thus, a
covering (\ref{covering2}) should result in violation of the Newtonian
cosmological macroscopic hypothesis (\ref{macro-cosmo}) and the
impossibility to determine a family of observes with physically reasonable
measurement devices which are able to perform proper space-time measurements
of the cosmological fluid characteristics.

The thorough analysis of the averaged fluid motion, physical foundations
underlying the definition of averages and the formalism for establishing
their properties, see Sections \ref{*stacf}, \ref{*rcaff} and \ref{*stamd},
has resulted in the derivation of the formulae (\ref{aver-material})-(\ref
{aver-material3}) for the space-time averaging out of the material
derivatives of cosmological fluid fields. It is especially important for
Newtonian cosmology where all these issues have a direct relation to the way
how our Universe is observed, how the cosmological data are measured and how
a theoretical model takes into account the measurement device
characteristics and the conditions of experiments. Using a covering
different than (\ref{covering}) and/or using the definitions of a
space-time, time or space averages different than (\ref{aver-space-time}),
(\ref{aver-time}) and (\ref{aver-space}) would lead to a violation of the
Newtonian cosmological macroscopic hypothesis (\ref{macro-cosmo}) and an
arrangement of a physically unreasonable set of measurement devices. That
would affect, as a result, the analytic properties of the averages to
abandon the Reynolds conditions (\ref{aver-linear})-(\ref{aver-idempotent})
and the formulae (\ref{aver-material})-(\ref{aver-material3}).

It is therefore important to analyze possible consequences of such different
choices in order to make a comparison with the approach presented here for
the formulation of the foundations of Newtonian cosmology and the averaged
Navier-Stokes-Poisson equations. Let us consider here a possibility to use
the Reynolds transport theorem (I-20) for derivation of the formula for a
space averaging out of the material derivatives. On the basis of the
Reynolds transport theorem is easy to show the following:

\begin{corollary}[The modified Reynolds transport theorem]
The rate of change of the ratio of the volume integral $F(t)$ (I-19),
\begin{equation}
F(t)=\int\limits_{\Sigma (t)}f(x^{i},t)dV,  \label{integral}
\end{equation}
of the cosmological fluid field $f(x^{i},t)$ over an arbitrary closed fluid
region $\Sigma (t)$ moving with the fluid to the value $V_{\Sigma }(t)$ of
the volume of the region (I-24),
\begin{equation}
\left\{ f\right\} (t)=\frac{F(t)}{V_{\Sigma }(t)}=\frac{1}{V_{\Sigma }(t)}
\int\limits_{\Sigma (t)}fdV,  \label{ratio}
\end{equation}
is given by
\begin{equation}
\frac{d\left\{ f\right\} (t)}{dt}=\left\{ \frac{df}{dt}\right\} (t)+\left\{ f
\frac{\partial u^{k}}{\partial x^{k}}\right\} (t)-\left\{ f\right\}
(t)\left\{ \frac{\partial u^{k}}{\partial x^{k}}\right\} (t).
\label{reynolds-mod}
\end{equation}
\end{corollary}

\noindent \textbf{Proof.}\hspace{0.4cm}The formula evidently follows from
the Reynolds transport theorem (I-20) and a corollary of the theorem (I-25)
giving the rate of change of the volume (I-24) of the region $\Sigma (t)$.
\hspace{0.4cm}\textbf{QED}\vspace{0.2cm}

The function of time $\left\{ f\right\} (t)$ measures the ratio (\ref{ratio}
) of the value of the quantity $f(x^{i},t)$ inside an arbitrary closed fluid
region $\Sigma (t)$ moving with the fluid to the value of the region's
volume. The modified Reynolds transport theorem (\ref{reynolds-mod})
determines the change of the ratio $\left\{ f\right\} (t)$ rather than of
the integral $F(t)$ (\ref{integral}) as the Reynolds transport theorem
(I-20) does and they are completely equivalent. All results in analysis of
the fluid kinematics made on the basis of (I-20) can be achieved by using
the modified Reynolds transport theorem (\ref{reynolds-mod}).

Though the quantity $\left\{ f\right\} (t)$ (\ref{ratio}) looks similar to a
space average $\left\langle f(x^{i},t)\right\rangle _{S}$ (\ref{aver-space})
of the cosmological fluid field $f(x^{i},t)$, it cannot be taken as a proper
definition of a space average of $f(x^{i},t)$ due to the following reasons:

(\#1) the quantity $\left\{ f\right\} (t)$, as well as the quantity $F(t),$
are well-defined functions of $t$ only, since they are defined only for
arbitrary closed fluid regions $\Sigma (t)$ moving with the fluid by the
meaning of the Reynolds transport theorem (I-20);

(\#2) the quantity $\left\{ f\right\} (t)$ cannot serve as a space average,
since no arrangements have been made for the construction of a fluid field $
\left\{ f\right\} (t)$ determined at each point $\{x^{i}\}$ and each moment
of time $t$, and, therefore, for an arrangement of a set of measurement
devices throughout the cosmological fluid configuration, see the First and
Second conditions of Section \ref{*stacf};

(\#3) the regions $\Sigma (t)$ moving with the fluid change their shape and
volume, which raises all the problems with (\ref{covering2}) discussed above;

(\#4) the Reynolds transport theorem (I-20) is known to be an integral
statement fully equivalent to the Euler expansion formula (I-15), see \cite
{Zala-AIC1:2002} and references therein, therefore any conclusions gained on
the basis of application of the modified Reynolds transport theorem (\ref
{reynolds-mod}) are completely equivalent to the corresponding statements
for the integrand of (\ref{integral}).

(\#5) the Reynolds transport theorem (I-20) and the modified Reynolds
theorem (\ref{reynolds-mod}) do not take into account by their physical
meaning and their conditions that the fluid particles of an averaged
cosmological fluid configuration have different equations of motion $
X^{i}=X^{i}(\xi ^{j},t)$ (\ref{flow-aver}), see Section \ref{*stamd}, as
compared with the equations of motion $x^{i}=x^{i}(\xi ^{j},t)$ (\ref{flow})
of the fluid particles of the corresponding inhomogeneous fluid
configuration and the region $\Sigma (t)$ is supposed to move along the
fluid particles' paths determined by (\ref{flow});

The modified Reynolds theorem (\ref{reynolds-mod}) cannot be therefore used
as a formula for a space averaging out of the material derivatives instead
of the formulae (\ref{aver-material})-(\ref{aver-material3}).

\section{The Ensemble Averages of a Cosmological Fluid}

\label{*eacf}

The use of space-time, time or space averages (\ref{aver-space-time}), (\ref
{aver-time}) and (\ref{aver-space}) is very convenient from the practical
point of view, but leads sometimes to analytical difficulties as it has been
discussed above. It is desirable therefore to find another method of
defining the average fluid fields such that it would have well-defined
analytic properties and would be universal. A convenient definition of this
type is known to be provided by the statistical picture of the fluid fields
as random fields (see \cite{Moni-Yagl:1971}, \cite{Moni-Yagl:1975}, \cite
{Lesl:1973}, \cite{Stan:1985}, \cite{Lin-Reid:1963}, \cite{FMRT:2001} and
reference therein), which has been established by Richardson \cite{Rich:1922}
, \cite{Rich:1926}, Taylor \cite{Tayl:1935}, Kolmogorov \cite{Kolm:1941},
\cite{Kol2:1941} and Kamp\'{e} de F\'{e}riet \cite{Kamp:1939}.

The basic feature of the statistical approach to the motion of fluid is the
transition from the consideration of a single moving fluid configuration
given by the equations of motion of fluid particles $x^{i}=x^{i}(\xi ^{j},t)$
(\ref{flow}) or $\xi ^{i}=\xi ^{i}(x^{j},t)$ (\ref{inverse-flow}) to the
consideration of the statistical ensemble of all similar moving fluid
configurations created by some set of fixed initial and boundary conditions.
If the values of the fluid velocity field $f(x^{i},t)$ for an evolving
inhomogeneous fluid configuration are measured many times under the same set
of conditions, the measured values $\left\langle f(x^{i},t)\right\rangle $
for the same moment of time $t$ and the same position $\{x^{i}\}$ are
expected to be different due to local spatial and temporal fluctuations of $
f(x^{i},t)$. However, the arithmetic mean of all these values at the moment
of time $t$ and the position $\{x^{i}\}$ can be expected to tend to a
definite value if a sufficiently large number of measurements has been
carried out. This arithmetic average over all possible realizations of the
fluid quantity $f(x^{j},t)$, that is, all the values $\left\langle
f(x^{j},t)\right\rangle $ as measured, in principle, by an observer, is
called an ensemble average of $f(x^{j},t)$ at the time $t$ and the position $
\{x^{i}\}$.\vspace{0.2cm}

\noindent \textbf{The ensemble average, or the probability mean, of a fluid
field }$f$
\begin{equation}
\left\langle f\right\rangle _{E}(x^{i},t)=\int\limits_{-\infty }^{+\infty
}fP_{(x^{i},t)}(f)df.  \label{aver-ensemble}
\end{equation}
\vspace{0.2cm}

\noindent Here the probability density function $P_{(x^{i},t)}(f)$ defined for all
times $t$ and points $\{x^{i}\}$ of the cosmological fluid configuration
determines the probabilities of different values $f(x^{i},t)$ among all
measured as
\begin{equation}
P_{(x^{i},t)}(f)df=\mathrm{Probabity\,}[f<f(x^{i},t)<f+df],\quad
\int\limits_{-\infty }^{+\infty }P_{(x^{i},t)}(f)df=1.
\label{probability-density}
\end{equation}
The quantity $\mathrm{Probabity\,}[f<f(x^{i},t)<f+df]$ in (\ref
{probability-density}) stands for the probability of $f(x^{i},t)$ to have
values between $f$ and $f+df$. A fluid field $f$ having a definite
probability density is called the random variable and the set of all
possible probabilities $P_{(x^{i},t)}(f^{\prime },f^{\prime \prime })=
\mathrm{Probabity\,}[f^{\prime }<f(x^{i},t)<f^{\prime \prime }]$
corresponding to $f$ is called its probability distribution.

The ensemble average (\ref{aver-ensemble}) is a well-defined function of a
reference moment $t$ of the absolute time, $t\in T$, and a reference
position $\{x^{k}\}$, $\{x^{k}\}\in S$, where the ensemble average $
\left\langle f\right\rangle _{E}(x^{i},t)$ is evaluated by a measurement
device with the measurement time interval $T$ and the space region $S\subset
E^{3}\ $determined by the measurement device resolution. It should be
pointed out here that the measured values of $f$ may be time, space, or
space-time averages in dependence on the physical conditions of a
cosmological fluid configuration under consideration. The nature of
measurements must be taken into account when making physical predictions and
interpretations on the basis of the statistical treatment of fluid dynamics.

Because the Navier-Stokes-Poisson equations are nonlinear one faces
necessity to deal with averaging products of the fluid fields. The ensemble
average (\ref{aver-ensemble}) for a product of cosmological fluid fields $
f=f(x^{i},t)$\ and $h=h(x^{i},t)$ defines the so-called ensemble correlation
function.\vspace{0.2cm}

\noindent \textbf{The two-point second order ensemble correlation, or
moment, function of fluid fields }$f$ \textbf{and }$h$\hspace{0.2cm}\emph{
The two-point second order correlation function is the ensemble average of a
product of }$f(x^{i},t)$\emph{\ and }$h(y^{j},s)$\textsc{\ }

\begin{equation}
\left\langle fh\right\rangle _{E}(x^{i},t;y^{j},s)=\int\limits_{-\infty
}^{+\infty }fhP_{(x^{i},t)(y^{j},s)}(f,h)dfdh.  \label{aver-ensemble-product}
\end{equation}
\vspace{0.2cm}

\noindent Here $P_{(x^{i},t)(y^{j},s)}(f,h)$ is the two-dimensional joint probability
density function defined for all times $t$ and $s$ and points $\{x^{i}\}$
and $\{y^{i}\}$ of the cosmological fluid configuration determines the
probabilities of different values $f(x^{i},t)$ and $h(y^{j},s)$ among all
measured as
\begin{gather}
P_{(x^{i},t)(y^{j},s)}(f,h)=\mathrm{Probabity\,}
[f<f(x^{i},t)<f+df,~h<h(y^{j},s)<h+dh],  \notag \\
\int\limits_{-\infty }^{+\infty }P_{(x^{i},t)(y^{j},s)}(f,h)dfdh=1.
\label{probability-density2}
\end{gather}
The quantity $\mathrm{Probabity\,}[f<f(x^{i},t)<f+df,~h<h(y^{j},s)<h+dh]$ in
(\ref{probability-density2}) stands for the probability of $f(x^{i},t)$ and $
h(y^{i},t)$ to have simultaneously values between $f$ and $f+df$ and between
$h$ and $h+dh$, correspondingly.

The correlation function (\ref{aver-ensemble-product}) is a well-defined
function of the reference moments $t$ and $s$ of the absolute time, $t,s\in
T $, and of the reference positions $\{x^{k}\}$ and $\{y^{k}\}$, $
\{x^{k}\},\{y^{k}\}\in S$, with the ensemble average $\left\langle
fh\right\rangle _{E}(x^{i},t;y^{j},s)$ being prescribed to the pair of the
Eulerian variables $\left( x^{i},t\right) $ and $(y^{j},s)$ and symmetric in
these variables,
\begin{equation}
\left\langle fh\right\rangle _{E}(x^{i},t;y^{j},s)_{E}=\left\langle
hf\right\rangle _{E}(y^{j},s;x^{i},t).  \label{aver-ensemble-product-sym}
\end{equation}

From the physical point of view the two-point ensemble average (\ref
{aver-ensemble-product}) corresponds to a simultaneous measurement of $f$
and $h$ by an observer with a measurement device with the time measurement
interval $T$ and the measurement device resolution region $S\subset E^{3}$.
As far as the Newtonian cosmological macroscopic hypothesis (\ref
{macro-cosmo}) is satisfied for a cosmological fluid configuration under
consideration a choice of the reference times $t$ and $s$ and the reference
positions $\{x^{k}\}$ and $\{y^{k}\}$ is arbitrary within $T$ and $S$.

The correlation functions of fluid fields are fundamental characteristic
functions responsible for essentially nonlinear phenomenon in evolving
fluids, such as, for example, turbulence and hydrodynamic instability \cite
{Reyn:1894}, \cite{Moni-Yagl:1971}, \cite{Kell-Frie:1924}, \cite{Tayl:1921}.
If some of such functions do not vanish for a fluid configuration, then the
corresponding fluid fields are said to have correlations.\vspace{0.2cm}

\noindent \textbf{The fluid field correlations}\hspace{0.2cm}\emph{A
cosmological fluid configuration has a central two-point second order
ensemble correlation, or moment, function }$C_{E}^{(2)}(x^{i},t;y^{j},s)$
\emph{, if there are, at least, two cosmological fluid fields }$f(x^{i},t)$
\emph{\ and }$h(x^{i},s)$\emph{\ such that}\textsc{\ }
\begin{equation}
C_{E}^{(2)}(x^{i},t;y^{j},s)=\left\langle fh\right\rangle
_{E}(x^{i},t;y^{j},s)-\left\langle f\right\rangle _{E}(x^{i},t)\left\langle
h\right\rangle _{E}(y^{j},s)\neq 0,  \label{func-ensemble-correlation}
\end{equation}
\begin{equation}
C_{E}^{(2)}(x^{i},t;y^{j},s)=C_{E}^{(2)}(y^{j},s;x^{i},t)
\label{func-ensemble-correlation-sym}
\end{equation}
\emph{in some open space region }$U\subset E^{3}$\emph{\ for an interval of
time }$\Delta t$, $x^{k},y^{k}\in U$, $t,s\in \Delta t$.\vspace{0.2cm}

One can define the central multi-point higher-order ensemble correlation
functions, $C_{E}^{(3)}(x^{i},t;y^{j},s;z^{k},r)$ and so on, by using
higher-dimensional joint probability density functions $
P_{(x^{i},t)(y^{j},s)(z^{j},r)}(f,h,g)$ and so on, whenever it is necessary
for analysis of the dynamics of a Newtonian universe by making definitions
similar to (\ref{aver-ensemble-product}) and (\ref{func-ensemble-correlation}
).

In order to characterize the ensemble averaged fluid configuration one needs
to determine now the ensemble averaged fields $\left\langle f\right\rangle
_{E}(x^{i},t)$ and the correlation function fields $\left\langle
fh\right\rangle _{E}(x^{i},t;y^{j},s)$ for each moments of time $t$ and $s$
at all positions $\{x^{k}\}$ and $\{y^{k}\}$. From the physical point of
view, that means that it is possible, at least in principle, to carry out
the measurements of the quantities$\ f$ and $h$ for all instants of time
during the evolution of a fluid configuration and at its each space
position. This requires a covering by the averaging time intervals $T$ and
the averaging space regions $S$ to be determined throughout the cosmological
fluid configuration in accordance with the First and Second conditions, see
Section \ref{*stacf}. It will be assumed that a covering (\ref{covering})
has been arranged throughout the cosmological fluid configuration and the
Newtonian cosmological macroscopic hypothesis (\ref{macro-cosmo}) is valid
for this set of measurement devices.

Thus one can formulate the main hypothesis on the statistical description of
Newtonian universes.\vspace{0.2cm}

\noindent \textbf{The Hypothesis of the statistical nature of Newtonian
universes: }\emph{The cosmological fluid fields of the components of the
fluid velocity vector }$u^{i}(x^{j},t)$\emph{, the fluid density }$\rho
(x^{ki},t)$\emph{, the fluid pressure }$p(x^{i},t)$ \emph{and the Newtonian
gravitational potential }$\phi (x^{i},t)$ \emph{are assumed to be random
fields defined for all times }$t$ \emph{and positions} $\{x^{i}\}$\emph{\ of
the cosmological fluid configuration with the corresponding probability
densities }$P_{(x^{j},t)}(u^{i})$, $P_{(x^{i},t)}(\rho )$, $P_{(x^{i},t)}(p)$
\emph{\ and} $P_{(x^{i},t)}(\phi )$. \emph{If the cosmological fluid fields
have correlations (\ref{func-ensemble-correlation}), that is, they are
statistically interconnected, it is assumed that there exist the joint
probability densities for corresponding fluid fields. }\vspace{0.2cm}

\section{The Properties of the Ensemble Averages}

The ensemble average (\ref{aver-ensemble}) satisfies the Reynolds conditions
(\ref{aver-linear})-(\ref{aver-idempotent}).

\begin{corollary}[The Reynolds conditions for the ensemble averages]
The ensemble average $\left\langle f\right\rangle
_{E}(x^{i},t)$ is a local single-valued function
\begin{equation}
\left( \frac{\partial }{\partial x^{i}}\frac{\partial }{\partial t}-\frac{
\partial }{\partial t}\frac{\partial }{\partial x^{i}}\right) \left\langle
f\right\rangle _{E}(x^{i},t)=0\,,\quad \left( \frac{\partial }{\partial x^{i}
}\frac{\partial }{\partial x^{j}}-\frac{\partial }{\partial x^{j}}\frac{
\partial }{\partial x^{i}}\right) \left\langle f\right\rangle
_{E}(x^{i},t)=0,  \label{ensemble-locality}
\end{equation}
of the reference time$\ t$ and position $\{x^{k}\}$ and it satisfies the
Reynolds conditions (\ref{aver-linear})-(\ref{aver-idempotent}):

(1*) the ensemble averaging is a linear operation
\begin{equation}
\left\langle af+bh\right\rangle _{E}(x^{i},t)=a\left\langle f\right\rangle
_{E}(x^{i},t)+b\left\langle f\right\rangle _{E}(x^{i},t),\quad \mathrm{if}
\quad a,b=\mathrm{const},  \label{aver-linear-ensemble}
\end{equation}

(2*) the ensemble averaging commutes with the partial differentiation
\begin{equation}
\frac{\partial }{\partial t}\left\langle f\right\rangle
_{E}(x^{i},t)=\left\langle \frac{\partial f}{\partial t}\right\rangle
_{E}(x^{i},t),\quad \frac{\partial }{\partial x^{i}}\left\langle
f\right\rangle _{E}(x^{i},t)=\left\langle \frac{\partial f}{\partial x^{i}}
\right\rangle _{E}(x^{i},t),  \label{aver-commutation-ensemble}
\end{equation}

(3*) the ensemble average is idempotent
\begin{equation}
\left\langle \left\langle f\right\rangle _{E}(x^{i},t)h\right\rangle
_{E}(y^{i},s)=\left\langle f\right\rangle _{E}(x^{i},t)\left\langle
h\right\rangle _{E}(y^{i},s)\quad \mathrm{or}\quad \left\langle \langle
f\rangle \right\rangle _{E}(x^{i},t)=\langle f\rangle _{E}(x^{i},t).
\label{aver-idempotent-ensemble}
\end{equation}
\end{corollary}

\noindent \textbf{Proof.}\hspace{0.4cm}The formulae
(\ref{ensemble-locality})-(\ref{aver-idempotent-ensemble}) follow
immediately from the definition of
the ensemble average (\ref{aver-ensemble}).\hspace{0.4cm}\textbf{QED}\vspace{
0.2cm}

The analytic properties of the ensemble averages are the same as those of
space-time, time and space averages. It is remarkable fact that they all are
rigorous properties for the ensemble averages. This is one of the
fundamental advantages of using this type of mean fluid fields.

The formula for the ensemble averaging out of the material derivatives can
be proved as follows.

\begin{theorem}[The ensemble average of the material derivative]
The ensemble average (\ref{aver-ensemble}) of the
material derivative of a cosmological fluid field $f(x^{i},t)$ is given by
the formula
\begin{equation}
\left\langle \frac{df}{dt}\right\rangle _{E}=\left\langle \frac{\partial f}{
\partial t}+u^{i}\frac{\partial f}{\partial x^{i}}\right\rangle _{E}=\frac{
d\left\langle f\right\rangle _{E}}{dt}+\left\langle u^{i}\frac{\partial f}{
\partial x^{i}}\right\rangle _{E}-\left\langle u^{i}\right\rangle _{E}\frac{
\partial \left\langle f\right\rangle _{E}}{\partial x^{i}}
\label{material-ensemble}
\end{equation}
where $\left\langle u^{i}f_{,i}\right\rangle _{E}=\left\langle
u^{i}f_{,i}\right\rangle _{E}(x^{j},t)$ is the single-point second order
correlation function (\ref{aver-ensemble-product}) of the fluid velocity $
u^{i}(x^{j},t)$ and the spatial derivative $\partial f(x^{i},t)/\partial
x^{i}$, and the material derivative of $\left\langle f\right\rangle
_{E}(x^{i},t)$ is
\begin{equation}
\frac{d\left\langle f\right\rangle _{E}}{dt}=\frac{\partial \left\langle
f\right\rangle _{E}}{\partial t}+\left\langle u^{i}\right\rangle \frac{
\partial \left\langle f\right\rangle _{E}}{\partial x^{i}}.
\label{material-aver-ensemble}
\end{equation}
\end{theorem}

\noindent \textbf{Proof.}\hspace{0.4cm}The proof is straightforward by
taking the ensemble average (\ref{aver-ensemble}) of the material derivative
(\ref{material-dr2}) of a cosmological fluid field $f=f(x^{i},t)$ with using
the Reynolds conditions (\ref{aver-commutation-ensemble}), the definitions
of the ensemble average $\left\langle f\right\rangle _{E}(x^{i},t)$ (\ref
{aver-ensemble}) of the cosmological fluid field and the ensemble average $
\left\langle u^{i}\right\rangle _{E}(x^{k},t)$ of the cosmological fluid
velocity $u^{i}(x^{j},t)$, and the introduction of the single-point second
order correlation function $\left\langle u^{i}f_{,i}\right\rangle
_{E}(x^{j},t)$ (\ref{aver-ensemble-product}).\hspace{0.4cm}\textbf{QED}
\vspace{0.2cm}

There are two equivalent useful forms of the formula (\ref{material-ensemble}
) which are completely analogous to the formulae (\ref{aver-material2}) and
(\ref{aver-material3}).

\begin{corollary}[The ensemble averaged Navier-Stokes field operator]
The ensemble average (\ref{aver-ensemble}) of the field
operator of the Navier-Stokes equation (II-10) has the following form:
\begin{equation}
\left\langle \frac{du^{i}}{dt}\right\rangle _{E}=\left\langle \frac{\partial
u^{i}}{\partial t}+u^{k}\frac{\partial u^{i}}{\partial x^{k}}\right\rangle
_{E}=\frac{d\left\langle u^{i}\right\rangle _{E}}{dt}+\left\langle u^{k}
\frac{\partial u^{i}}{\partial x^{k}}\right\rangle _{E}-\left\langle
u^{k}\right\rangle _{E}\frac{\partial \left\langle u^{i}\right\rangle _{E}}{
\partial x^{k}}  \label{navier-stokes-aver-ensemble}
\end{equation}
where $\left\langle u^{j}u_{,j}^{i}\right\rangle _{E}=\left\langle
u^{j}u_{,j}^{i}\right\rangle _{E}(x^{k},t)$ is the single-point second order
correlation function (\ref{aver-ensemble-product}) of the fluid velocity $
u^{i}(x^{j},t)$ and its spatial derivative $\partial u^{i}(x^{k},t)/\partial
x^{j}$ and the material derivative of the ensemble averaged fluid velocity $
\left\langle u^{i}\right\rangle _{E}(x^{j},t)$ is given by (\ref
{material-aver-ensemble}).
\end{corollary}

One can now establish the analytic properties of the ensemble correlation
function $\left\langle fh\right\rangle _{E}(x^{i},t;y^{j},s)$ (\ref
{aver-ensemble-product}) and prove the properties analogous to the
conditions (\ref{aver-commutation-corr})-(\ref{aver-idempotent-corr}).

\begin{corollary}[The properties of the ensemble correlation functions]
The two-po- int second order moment function (\ref{aver-ensemble-product}), $
\left\langle fh\right\rangle _{E}(x^{i},t;y^{j},s)$, of two cosmological
fluid fields $f(x^{i},t)$\ and $h(x^{i},t)$ has the following properties:

(1*) it is a bilocal single-valued function of the reference times$\ $and
positions, $(x^{k},t)$ and $(y^{k},s)$, such as all second antisymmetrized derivatives
with respect to all pairs of the variables $(x^{k},t;y^{k},s)$ vanish;

(2*) the two-point second order moment function $\left\langle
fh\right\rangle _{E}(x^{i},t;y^{j},s)$ satisfies the conditions of the
partial differentiation
\begin{align}
\frac{\partial }{\partial x^{k}}\left\langle fh\right\rangle
_{E}(x^{i},t;y^{j},s)& =\left\langle \frac{\partial f}{\partial x^{k}}
h\right\rangle _{E}(x^{i},t;y^{j},s),  \notag \\
\frac{\partial }{\partial t}\left\langle fh\right\rangle
_{E}(x^{i},t;y^{j},s)& =\left\langle \frac{\partial f}{\partial t}
h\right\rangle _{EE}(x^{i},t;y^{j},s),\quad \mathrm{etc}
\label{aver-com-cor-ensemble}
\end{align}
for all variables $(x^{k},t;y^{k},s)$;

(3*) the two-point second order ensemble moment function $\left\langle
fh\right\rangle _{E}(x^{i},t;y^{j},s)$ is idempotent
\begin{equation}
\left\langle \left\langle fh\right\rangle
_{E}(x^{i},t;y^{j},s)g\right\rangle _{E}(z^{k},r)=\left\langle
fh\right\rangle _{E}(x^{i},t;y^{j},s)\left\langle g\right\rangle
_{E}(z^{k},r)\quad \mathrm{or}  \label{aver-idem-corr0-ensemble}
\end{equation}
\begin{equation}
\left\langle \left\langle fh\right\rangle \right\rangle
_{E}(x^{i},t;y^{j},s)=\left\langle fh\right\rangle _{E}(x^{i},t;y^{j},s)
\label{aver-idem-corr-ensemble}
\end{equation}
where $g=g(z^{k},r)$ is another cosmological fluid field.
\end{corollary}

\noindent \textbf{Proof.}\hspace{0.4cm}The formulae (\ref
{aver-com-cor-ensemble})-(\ref{aver-idem-corr-ensemble}) follow immediately
from the definition of the ensemble correlation function (\ref
{aver-ensemble-product}).\hspace{0.4cm}\textbf{QED}\vspace{0.2cm}

Thus, one can work with the ensemble correlation functions as with usual
local functions of many variables depending on the reference times$\ $and
positions by differentiating and integrating them, expanding into series,
solving differential and integral equations for them using standard
techniques of mathematical physics.

It should be noted here that the conditions (\ref{aver-com-cor-ensemble})
mean that a partial derivative of an ensemble correlation function produces
an ensemble correlation function of the same order including the
corresponding partial derivative of a fluid field.

There is an important asymptotic property \cite{Moni-Yagl:1971} of the
central ensemble correlation function $C_{E}^{(2)}(x^{i},t;y^{j},s)$ (\ref
{func-ensemble-correlation}), which is of particular interest in Newtonian
cosmology.\vspace{0.2cm}

\noindent \textbf{The asymptotic condition for the central two-point second
order ensemble correlation function}\textsc{\hspace{0.2cm}}\emph{For any two
cosmological fluid fields }$f(x^{i},t)$\emph{\ and }$h(x^{i},t)$\emph{\
which are correlated (\ref{func-ensemble-correlation}) and satisfy the
Newtonian cosmological macroscopic hypothesis (\ref{macro-cosmo}), their
central two-point second order moment function }$
C_{E}^{(2)}(x^{i},t;y^{j},s) $\emph{\ tends to zero when the distance
between the points }$\{x^{i}\}$\emph{\ and }$\{y^{i}\}$\emph{\ and the time
interval between }$t$\emph{\ and }$s$\emph{\ infinitely grow}\textsc{\ }
\begin{equation}
C_{E}^{(2)}(x^{i},t;y^{j},s)\rightarrow 0\quad \mathrm{as}\quad \left[
\delta _{ij}(x^{i}-y^{i})(x^{j}-y^{j})\right] ^{1/2}\rightarrow \infty
,\quad \mid s-t\mid \rightarrow \infty .
\label{correlation-asymptotic-ensemble}
\end{equation}
\vspace{0.2cm}

\section{The Ergodic Hypothesis of Newtonian Cosmology}

\label{*ehmc}

The ensemble averages are undoubtedly more convenient for analytic
description of moving fluids. Adopting a hypothesis on the existence of the
probability distributions for all fluid fields, such the Hypothesis of the
statistical nature of Newtonian universes in Newtonian cosmology, see
Section \ref{*eacf}, one can use the mathematical techniques of the modern
probability theory and statistical mechanics. The ensemble averages (\ref
{aver-ensemble}) and (\ref{aver-ensemble-product}) are defined uniquely (\ref
{ensemble-locality}) throughout a cosmological fluid configuration and have
all properties naturally required, such as the Reynolds conditions (\ref
{aver-linear-ensemble})-(\ref{aver-idempotent-ensemble}) and the conditions
(\ref{aver-com-cor-ensemble})-(\ref{aver-idem-corr-ensemble}). It is
important, however, to note that in the framework of this approach, an
important new question arises regarding the comparison of the theoretical
deductions following from the statistical treatment of the dynamics of
fluids with the data of direct space-time, time and space measurements.

According to its definition (\ref{aver-ensemble}) the ensemble average is a
mean value taken over all possible values of the fluid field under study,
which are measured by a measurement device with a time measurement interval
$T$ and a measurement space region $S$ determined by the device resolution.
Thus, to determine the average values by an experiment or, say, by
cosmological observations, with an acceptable high accuracy, one should
carry out a large number of measurements in many set of repeated similar
observations or experiments. In practice, however, no such multiply repeated
measurements are available, or even possible to perform. Therefore in most
cases one is forced to determine the average values from the data taken in
the course of a single experiment where the data obtained either by a
space-time, time or space averaging over a time interval $T$ and/or a space
region $S$. From the physical point of view it means that the assumption
about the existence of the probability distributions, see the Hypothesis of
the statistical nature of Newtonian universes, see Section \ref{*eacf}, does
not by itself eliminate the problem of the validity of using the space-time
(\ref{aver-space-time}), time (\ref{aver-time}) or space averages (\ref
{aver-space}) for description and analysis of moving inhomogeneous fluids.
But in this context the problem has acquired a different formulation and
interpretation. Indeed, instead of investigating the particular properties
of these averaging procedures, one should establish how close the
space-time, time or space averaged values of the fluid fields measured in
real observations and experiments are to the corresponding ensemble average
values. It is known from the statistical mechanics that a replacement of an
ensemble average value of a random quantity calculated over all possible
states of this quantity by the directly measurable time average of the same
quantity may be only possible if the averaging time interval $T$ becomes
infinitely large. Under this condition the time average converges to the
corresponding ensemble average \cite{Moni-Yagl:1971}, \cite{Moni-Yagl:1975},
\cite{Lesl:1973}, \cite{Stan:1985}, \cite{Lin-Reid:1963}, \cite{FMRT:2001}.
In certain very special cases, the validity of this assumption may be proved
rigorously. In the most cases of interest this fundamental physical
hypothesis of the consistency between theoretical models, their experimental
verification and the real physical phenomenon remains unproved and it is
usually adopted as an additional hypothesis called the Ergodic hypothesis.
It is formulated here for Newtonian cosmology.\vspace{0.2cm}

\noindent \textbf{The Ergodic hypothesis of Newtonian cosmology: }\emph{If a
cosmological fluid configuration satisfies the Newtonian cosmological
macroscopic hypothesis (\ref{macro-cosmo}) and the Hypothesis of the
statistical nature of Newtonian universes, see Section \ref{*eacf}, and a
covering (\ref{covering}) by the averaging time intervals }$T$\emph{\ and
the space regions }$S$\emph{\ is determined throughout the cosmological
fluid configuration, then for any cosmological fluid field }$f(x^{i},t)$
\emph{\ the following conditions are assumed to hold:}
\vspace{0.2cm}

\noindent \emph{(1) its ensemble average }$\left\langle f\right\rangle _{E}(x^{i},t)$
\emph{(\ref{aver-ensemble}) is equal to its time average }$\left\langle
f(x^{i},t)\right\rangle _{T}$\emph{\ (\ref{aver-time}) over an infinitely
large time interval }$T$\emph{\ and both averages do not depend on time }$t$
(\ref{aver-time-stationary}),
\begin{equation}
\lim_{T\rightarrow \infty }\left\langle f(x^{i},t)\right\rangle
_{T}=\left\langle f(x^{i})\right\rangle _{T}=\left\langle f\right\rangle
_{E}(x^{i}),  \label{ergodic-time}
\end{equation}
\vspace{0.2cm}

\noindent \emph{(2) its ensemble average }$\left\langle f\right\rangle _{E}(x^{i},t)$
\emph{(\ref{aver-ensemble}) is equal to its space average }$\left\langle
f(x^{i},t)\right\rangle _{S}$\emph{\ (\ref{aver-space}) over the whole space
}$S=E^{3}$ \emph{and both averages do not depend on a position }$\{x^{i}\}$
\emph{\ }(\ref{aver-space-stationary}),
\begin{equation}
\lim_{V_{S}\rightarrow \infty }\left\langle f(x^{i},t)\right\rangle
_{S}=\left\langle f(t)\right\rangle _{S}=\left\langle f\right\rangle _{E}(t),
\label{ergodic-space}
\end{equation}
\vspace{0.2cm}

\noindent \emph{(3) its ensemble average }$\left\langle f\right\rangle _{E}(x^{i},t)$
\emph{(\ref{aver-ensemble}) is equal to its space-time average }$\left\langle
f(x^{i},t)\right\rangle _{ST}$\emph{\ (\ref{aver-space-time}) over an
infinitely large time interval }$T$\emph{\ and over the whole space }$
S=E^{3} $ \emph{and both averages do not depend on on time }$t$\emph{\ and a
position }$\{x^{i}\}$\emph{\ (\ref{aver-time-stationary}) and (\ref
{aver-space-stationary}),}
\begin{equation}
\lim_{T\rightarrow \infty }\lim_{V_{S}\rightarrow \infty }\left\langle
f(x^{i},t)\right\rangle _{ST}=\left\langle f\right\rangle _{ST}=\left\langle
f\right\rangle _{E}.  \label{ergodic-space-time}
\end{equation}
\vspace{0.2cm}

Thus, if proved, the Ergodic hypothesis of Newtonian cosmology (\ref
{ergodic-time}), (\ref{ergodic-space}) and (\ref{ergodic-space-time})
guarantees that the values of cosmological fluid quantities measured by an
observer in a Newtonian universe by means of space-time, time and/or space
averagings are equal within the accuracy of measurements to the values
predicted theoretically on the basis of the statistical treatment of the
evolution of the Newtonian universe.

\section{The Averaged Navier-Stokes-Poisson Equations in Kinematic Quantities
}

The system of the averaged Navier-Stokes-Poisson equations in kinematic
quantities can be now established. Since the space-time (\ref
{aver-space-time}) and ensemble averaging, (\ref{aver-space-time}) and (\ref
{aver-ensemble}), have the same analytic properties (\ref{aver-locality})
and (\ref{ensemble-locality}), satisfy the Reynolds conditions (\ref
{aver-commutation})-(\ref{aver-idempotent}) and (\ref{aver-linear-ensemble}
)-(\ref{aver-idempotent-ensemble}), and the same formulae for averaging out
the material derivatives (\ref{aver-material}), (\ref{navier-stokes-aver})
and (\ref{material-ensemble}), (\ref{navier-stokes-aver-ensemble}), the
notation $\left\langle f\right\rangle (x^{i},t)$ will used hereafter for
both averaging procedures. Both procedures lead to the same form of the
averaged equations, though they are physically equivalent, strictly
speaking, only under the Ergodic hypothesis of Newtonian cosmology (\ref
{ergodic-space-time}), see Section \ref{*ehmc}. Use of the time and space
averages (\ref{aver-time}), (\ref{aver-space}), see Sections \ref{*tacf} and
\ref{*sacf}, can be made every time when the physical conditions of a
Newtonian universe require this and the corresponding parts of the Ergodic
hypothesis of Newtonian cosmology, (\ref{ergodic-time}) and (\ref
{ergodic-space}), are valid.

\begin{theorem}[The averaged Navier-Stokes-Poisson equations]
If a covering (\ref{covering}) by the averaging time intervals $T$ and the
space regions $S$ is determined through the cosmological fluid
configuration, then the averaged system of the Navier-Stokes-Poisson
equations in terms of the kinematic quantities (II-4), (II-6), (II-8),
(II-12), (II-51)-(II-56), (II-59)-(II-61), (I-64)-(II-66), (II-69), (II-28)
and (II-70) is as follows.
\end{theorem}
\vspace{0.2cm}

\noindent \textbf{The averaged Raychaudhuri evolution equation for the
averaged expansion scalar }$\left\langle \theta \right\rangle $
\begin{equation}
\frac{d\left\langle \theta \right\rangle }{dt}+\left\langle u^{i}\frac{
\partial \theta }{\partial x^{i}}\right\rangle -\left\langle
u^{i}\right\rangle \frac{\partial \left\langle \theta \right\rangle }{
\partial x^{i}}+\frac{1}{3}\left\langle \theta ^{2}\right\rangle
+2(\left\langle \sigma ^{2}\right\rangle -\left\langle \omega
^{2}\right\rangle )+4\pi G\left\langle \rho \right\rangle -\Lambda -\delta
^{ij}\left\langle A_{i}\right\rangle _{,j}=0,  \label{evol-expansion-aver}
\end{equation}
\vspace{0.2cm}

\noindent \textbf{The averaged propagation equation for the averaged shear
tensor }$\left\langle \sigma _{ij}\right\rangle $
\begin{gather}
\frac{d\left\langle \sigma _{ij}\right\rangle }{dt}+\left\langle u^{k}\frac{
\partial \sigma _{ij}}{\partial x^{k}}\right\rangle -\left\langle
u^{k}\right\rangle \frac{\partial \left\langle \sigma _{ij}\right\rangle }{
\partial x^{k}}+\delta ^{kl}\left\langle \sigma _{ik}\sigma
_{lj}\right\rangle +\frac{2}{3}\left\langle \theta \sigma _{ij}\right\rangle
-  \notag \\
\frac{1}{3}\delta _{ij}(2\left\langle \sigma ^{2}\right\rangle +\left\langle
\omega ^{2}\right\rangle -\delta ^{kl}\left\langle A_{k}\right\rangle
_{,l})+\left\langle \omega _{i}\omega _{j}\right\rangle +\left\langle
E_{ij}\right\rangle -\left\langle A_{(i}\right\rangle _{,j)}=0,
\label{evol-shear-aver}
\end{gather}
\vspace{0.2cm}

\noindent \textbf{The averaged propagation equation for the averaged
vorticity vector }$\left\langle \omega ^{i}\right\rangle $
\begin{equation}
\frac{d\left\langle \omega ^{i}\right\rangle }{dt}+\left\langle u^{j}\frac{
\partial \omega ^{i}}{\partial x^{j}}\right\rangle -\left\langle
u^{j}\right\rangle \frac{\partial \left\langle \omega ^{i}\right\rangle }{
\partial x^{j}}+\frac{2}{3}\left\langle \theta \omega ^{i}\right\rangle
-\delta ^{ij}\left\langle \sigma _{jk}\omega ^{k}\right\rangle -\frac{1}{2}
\varepsilon ^{ijk}\left\langle A_{j}\right\rangle _{,k}=0,
\label{evol-vorticity-aver}
\end{equation}
\vspace{0.2cm}

\noindent \textbf{The averaged kinematic decomposition of the averaged
tensor }$\left\langle u_{i,j}\right\rangle $
\begin{equation}
\left\langle u_{i}\right\rangle _{,j}=\left\langle \sigma _{ij}\right\rangle
+\frac{1}{3}\delta _{ij}\left\langle \theta \right\rangle +\left\langle
\omega _{ij}\right\rangle ,\hspace{0.4cm}\left\langle \sigma
_{ij}\right\rangle =\left\langle \sigma _{ji}\right\rangle ,\hspace{0.4cm}
\left\langle \theta \right\rangle =\delta ^{ij}\left\langle \sigma
_{ij}\right\rangle ,\hspace{0.4cm}\left\langle \omega _{ij}\right\rangle
=-\left\langle \omega _{ji}\right\rangle ,  \label{decomposition2-aver}
\end{equation}
\vspace{0.2cm}

\noindent \textbf{The averaged first identity for the averaged tensor }$
\left\langle u_{i,j}\right\rangle $\textsc{\ }
\begin{equation}
\left\langle u_{i}\right\rangle _{,jt}=\left\langle u_{i}\right\rangle
_{,tj},  \label{identity1-aver}
\end{equation}
\vspace{0.2cm}

\noindent \textbf{The averaged second identity for the averaged tensor }$
\left\langle u_{i,j}\right\rangle $
\begin{equation}
\left\langle u_{i}\right\rangle _{,jk}=\left\langle u_{i}\right\rangle
_{,kj},  \label{identity2-aver}
\end{equation}
\vspace{0.2cm}

\noindent \textbf{The averaged first constraint equation}
\begin{equation}
\delta ^{jk}(\left\langle \sigma _{ij}\right\rangle _{,k}-\left\langle
\omega _{ij}\right\rangle _{,k})-\frac{2}{3}\left\langle \theta
\right\rangle _{,j}=0,  \label{constraint-1-aver}
\end{equation}
\vspace{0.2cm}

\noindent \textbf{The averaged second constraint equation}\textsc{\ }
\begin{equation}
\left\langle \omega ^{i}\right\rangle _{,i}=0,  \label{constraint-2-aver}
\end{equation}
\vspace{0.2cm}

\noindent \textbf{The averaged third constraint equation}
\begin{equation}
\delta ^{j(i}(\left\langle \sigma _{jk}\right\rangle _{,l}+\left\langle
\omega _{jk}\right\rangle _{,l})\varepsilon ^{m)kl}=0,
\label{constraint-3-aver}
\end{equation}
\vspace{0.2cm}

\noindent \textbf{The averaged first integrability condition}\textsc{\ }
\begin{equation}
\left\langle E_{i}^{j}\right\rangle _{,j}=\frac{8\pi G}{3}\left\langle \rho
\right\rangle _{,i},  \label{integrability-1-aver}
\end{equation}
\vspace{0.2cm}

\noindent \textbf{The averaged second integrability condition}
\begin{equation}
\left\langle E_{k}^{(i}\right\rangle _{,j}\varepsilon ^{l)kj}=0,
\label{integrability-2-aver}
\end{equation}
\vspace{0.2cm}

\noindent \textbf{The averaged third integrability condition}
\begin{equation}
\left\langle \sigma _{\lbrack k}^{[i}\right\rangle _{,l]}^{,j]}+\frac{2}{3}
\delta _{\lbrack k}^{[i}\left\langle \theta \right\rangle _{,l]}^{,j]}=0,
\label{integrability-3-aver}
\end{equation}
\vspace{0.2cm}

\noindent \textbf{The averaged equation of continuity as the evolution
equation for the averaged fluid density }$\left\langle \rho \right\rangle $
\begin{equation}
\frac{d\left\langle \rho \right\rangle }{dt}+\left\langle u^{i}\frac{
\partial \rho }{\partial x^{i}}\right\rangle -\left\langle
u^{i}\right\rangle \frac{\partial \left\langle \rho \right\rangle }{\partial
x^{i}}+\left\langle \rho \theta \right\rangle =0,  \label{evol-density-aver}
\end{equation}
\vspace{0.2cm}

\noindent \textbf{The averaged equation of state}
\begin{equation}
\left\langle \rho \right\rangle =\left\langle \rho (p)\right\rangle ,
\label{eq-state-aver}
\end{equation}
\vspace{0.2cm}

\noindent \textbf{The averaged total acceleration }$\left\langle
A_{i}\right\rangle $\textbf{\ by the averaged Navier-Stokes equation}\textsc{
\ }
\begin{equation}
\left\langle A_{i}\right\rangle =-\left\langle \frac{1}{\rho }
p_{,i}\right\rangle ,  \label{navier-stokes-total-aver}
\end{equation}
\vspace{0.2cm}

\noindent \textbf{The averaged Poisson equation for the averaged Newtonian
gravitational potential }$\left\langle \phi \right\rangle $
\begin{equation}
\delta ^{ij}\left\langle g_{i}\right\rangle _{,j}=4\pi G\left\langle \rho
\right\rangle -\Lambda ,\hspace{0.4cm}\mathrm{or}\hspace{0.4cm}\delta
^{ij}\left\langle \phi \right\rangle _{,ij}=4\pi G\left\langle \rho
\right\rangle -\Lambda ,  \label{poisson-aver}
\end{equation}
\vspace{0.2cm}

\noindent \textbf{The averaged identity for the averaged Newtonian
gravitational acceleration }$\left\langle g_{i}\right\rangle $
\begin{equation}
\varepsilon ^{ijk}\left\langle g_{j}\right\rangle _{,k}=0,
\label{identity-aver}
\end{equation}
\vspace{0.2cm}

\noindent \textbf{The averaged tidal force tensor} $\left\langle
E_{ij}\right\rangle $ \textbf{for the averaged Newtonian gravitational
potential }$\left\langle \phi \right\rangle $

\begin{equation}
\left\langle E_{ij}\right\rangle =\left\langle \phi \right\rangle _{,ij}-
\frac{1}{3}\delta _{ij}\delta ^{kl}\left\langle \phi \right\rangle _{,kl},
\hspace{0.4cm}\left\langle E\right\rangle _{ij}=\left\langle
E_{ji}\right\rangle ,\hspace{0.4cm}\delta ^{kl}\left\langle
E_{kl}\right\rangle =0,  \label{curvature-newton}
\end{equation}
\vspace{0.2cm}

\noindent \textbf{The averaged evolution equations for the averaged tidal
force tensor }$\left\langle E_{ij}\right\rangle $
\begin{gather}
\frac{d\left\langle E_{ij}\right\rangle }{dt}+\left\langle u^{k}\frac{
\partial E_{ij}}{\partial x^{k}}\right\rangle -\left\langle
u^{k}\right\rangle \frac{\partial \left\langle E_{ij}\right\rangle }{
\partial x^{k}}+\left\langle \theta E_{ij}\right\rangle -\delta _{ij}\delta
^{kn}\delta ^{lm}\left\langle \sigma _{kl}E_{nm}\right\rangle -  \notag \\
3\delta ^{kl}\left\langle \sigma _{k(i}E_{j)l}\right\rangle -\delta
^{kl}\left\langle \omega _{k(i}E_{j)l}\right\rangle +4\pi G\left\langle \rho
\sigma _{ij}\right\rangle =0,  \label{evol-tidal-aver}
\end{gather}
\vspace{0.2cm}

\noindent \textbf{Proof.}\hspace{0.4cm}The averaging out of the system of
the Navier-Stokes-Poisson equations in terms of kinematic quantities (II-4),
(II-6), (II-8), (II-12), (II-51)-(II-56), (II-59)-(II-61), (I-64)-(II-66),
(II-69), (II-28) and (II-70) is straightforward by applying the Reynold
conditions (\ref{aver-commutation})-(\ref{aver-idempotent}) or (\ref
{aver-linear-ensemble})-(\ref{aver-idempotent-ensemble}), and the formulae
for averaging out the material derivatives (\ref{aver-material}), (\ref
{navier-stokes-aver}) or (\ref{material-ensemble}), (\ref
{navier-stokes-aver-ensemble}).\hspace{0.4cm}\textbf{QED}\vspace{0.2cm}

The system of the averaged Navier-Stokes-Poisson (\ref{evol-expansion-aver}
)-(\ref{evol-tidal-aver}) contains the following independent set of the
single-point second order moment functions:\vspace{0.2cm}

\newpage

\noindent \textbf{The single-point second order moments in the averaged
Navier-Stokes-Poisson equations}: \vspace{0.2cm}

\noindent \emph{The fluid velocity }$u^{i}$\emph{\ with the spatial
derivatives of the kinematic quantities }$\sigma _{ij}$\emph{, }$\theta $
\emph{\ and }$\omega ^{i}$\emph{, the fluid density }$\rho $\emph{\ and the
tidal force tensor }$E_{ij}$

\begin{equation}
\left\langle u^{i}\frac{\partial \theta }{\partial x^{i}}\right\rangle ,
\hspace{0.4cm}\left\langle u^{k}\frac{\partial \sigma _{ij}}{\partial x^{k}}
\right\rangle ,\hspace{0.4cm}\left\langle u^{j}\frac{\partial \omega ^{i}}{
\partial x^{j}}\right\rangle ,\hspace{0.4cm}\left\langle u^{i}\frac{\partial
\rho }{\partial x^{i}}\right\rangle ,\hspace{0.4cm}\left\langle u^{k}\frac{
\partial E_{ij}}{\partial x^{k}}\right\rangle ,  \label{moment-velocity}
\end{equation}
\vspace{0.2cm}

\noindent \emph{The kinematic quantities }$\sigma _{ij}$\emph{, }$\theta $
\emph{\ and }$\omega ^{i}$\emph{\ themselves}
\begin{equation}
\left\langle \theta ^{2}\right\rangle ,\hspace{0.4cm}\delta
^{kl}\left\langle \sigma _{ik}\sigma _{lj}\right\rangle ,\hspace{0.4cm}
\left\langle \theta \sigma _{ij}\right\rangle ,\hspace{0.4cm}\left\langle
\omega _{i}\omega _{j}\right\rangle ,\hspace{0.4cm}\left\langle \theta
\omega ^{i}\right\rangle ,\hspace{0.4cm}\delta ^{ij}\left\langle \sigma
_{jk}\omega ^{k}\right\rangle ,  \label{moment-kinematic}
\end{equation}
\vspace{0.2cm}

\noindent \emph{The fluid density }$\rho $ \emph{with the kinematic
quantities }$\sigma _{ij}$\emph{\ and }$\theta $\emph{\ and the pressure
gradient }$p_{,i}$
\begin{equation}
\left\langle \rho \theta \right\rangle ,\hspace{0.4cm}\left\langle \rho
\sigma _{ij}\right\rangle ,\hspace{0.4cm}\left\langle \frac{1}{\rho }
p_{,i}\right\rangle ,  \label{moment-density}
\end{equation}
\vspace{0.2cm}

\noindent \emph{The tidal force tensor }$E_{ij}$ \emph{with the kinematic
quantities }$\sigma _{ij}$\emph{, }$\theta $\emph{\ and }$\omega ^{i}$
\begin{equation}
\left\langle \theta E_{ij}\right\rangle ,\hspace{0.4cm}\delta _{ij}\delta
^{km}\delta ^{lm}\left\langle \sigma _{kl}E_{mn}\right\rangle ,\hspace{0.4cm}
\delta ^{kl}\left\langle \sigma _{k(i}E_{j)l}\right\rangle ,\hspace{0.4cm}
\delta ^{kl}\left\langle \omega _{k(i}E_{j)l}\right\rangle .
\label{moment-tidal}
\end{equation}
\vspace{0.2cm}

The system of the averaged Navier-Stokes-Poisson (\ref{evol-expansion-aver}
)-(\ref{evol-tidal-aver}) is not closed with respect to these single-point
second order moment functions. The system should be supplemented by the
equations for these moment functions (\ref{moment-velocity})-(\ref
{moment-tidal}). A derivation of a system of the equations for the second
moments is known \cite{Reyn:1894}-\cite{Moni-Yagl:1975}, \cite{Lesl:1973},
\cite{Kell-Frie:1924}, \cite{Stan:1985}, \cite{Lin-Reid:1963}, \cite
{FMRT:2001} to produce the higher-order moments of fluid fields. This is the
so-called closure problem. To make an analysis possible and also take into
account properties of particular fluid configurations one must formulate
some conditions for the fluid field moments of some order to terminate this
infinite set of equations.

\section*{Acknowledgments}

Roustam Zalaletdinov would like to thank Alan Coley for his hospitality in
Dalhousie University and acknowledges partial support of this work by the
Swiss National Science Foundation, Grant 7BYPJ065731.

\end{document}